\documentclass[aps,prd,reprint,groupedaddress,amsmath,amssymb,floatfix]{revtex4-2}

\usepackage{graphicx} 
\usepackage{dcolumn}  
\usepackage{bm}       
\usepackage{hyperref} 
\usepackage{mathrsfs} 
\usepackage{color}

\begin{document}

\title{Charmed baryon-to-meson ratios inside jets as a discriminator of hadronization mechanisms: from color reconnection in p+p to coalescence in Pb+Pb}

\author{Zi-Xuan Xu}
\affiliation{Key Laboratory of Quark \& Lepton Physics (MOE) and Institute of Particle Physics, Central China Normal University, Wuhan 430079, China}

\author{Wei Dai}
\email{corresponding author: weidai@cug.edu.cn}
\affiliation{School of Mathematics and Physics, China University of Geosciences, Wuhan 430074, China}

\author{Ben-Wei Zhang}
\affiliation{Key Laboratory of Quark \& Lepton Physics (MOE) and Institute of Particle Physics, Central China Normal University, Wuhan 430079, China}
\affiliation{Guangdong Provincial Key Laboratory of Nuclear Science, Institute of Quantum Matter, South China Normal University, Guangzhou 510006, China}

\author{Enke Wang}
\affiliation{State Key Laboratory of Nuclear Physics and Technology, Institute of Quantum Matter, South China Normal University, Guangzhou 510006, China}

\affiliation{Guangdong Basic Research Center of Excellence for Structure and Fundamental Interactions of Matter, Guangdong Provincial Key Laboratory of Nuclear Science, Guangzhou 510006, China}

\date{\today}

\begin{abstract}
Jets separate the products of hard scatterings from the underlying event and thus help to clarify the origin of the enhanced baryon-to-meson ratios observed from small to large collision systems. We investigate the charmed baryon-to-meson ratio $\Lambda_c^+/D^0$ within jets in p+p and Pb+Pb collisions at $\sqrt{s_{NN}}=5.02$~TeV to discriminate between candidate hadronization mechanisms. In p+p collisions, simulated with PYTHIA8 using color reconnection (CR-BLC Mode 2) as a representative beyond-leading-color scenario, the in-jet ratio is enhanced at low-to-intermediate $p_T$ relative to the inclusive one---opposite to the light-flavor $\Lambda/K^0_S$ case---because the color-reconnection-induced excess of the ratio remains correlated with the jet and survives the jet requirement, while the fragmentation baseline shows no such enhancement. In Pb+Pb collisions, described by the SHELL model with a coalescence plus fragmentation hybrid hadronization, the in-jet ratio is instead suppressed moderately at intermediate $p_T$, because coalescence hadrons assembled from soft thermal partons carry no correlation with the jet direction and are preferentially discarded by the jet selection. The sign of the in-jet modification thus discriminates between the two mechanisms. Both mechanisms shift charmed hadrons toward small values of the parallel jet momentum fraction $z^{ch}_{||}$, with a stronger effect on $\Lambda_c^+$, so that the $z^{ch}_{||}$ spectrum probes hadronization mechanisms independently of parton energy loss. The radial profile of the in-jet ratio rises with the distance from the jet axis in both systems, with the lowest-$p_T$ interval displaced farthest for color reconnection and the intermediate-$p_T$ interval for coalescence. These results provide the baseline and predictions needed for future jet-tagged charmed-hadron measurements in both collision systems.
\end{abstract}


\maketitle

\section{Introduction}
An enhancement of the baryon-to-meson yield ratio at intermediate transverse momentum $p_T$, first established in nucleus--nucleus collisions, has since been observed also in high-multiplicity p+p and p+A collisions, where it exhibits a similar pronounced maximum~\cite{STAR:2007zea,STAR:2010yyv,STAR:2011iap,ALICE:2017thy,ALICE:2021vxl,CMS:2019uws,CMS:2024tjo,ALICE:2020wfu,ALICE:2020wla,PHENIX:2003tvk,STAR:2019ank,ALICE:2015dtd,ALICE:2021bib,ALICE:2019hno}.
Considerable effort has been devoted to understanding this baryon enhancement~\cite{Fries:2003vb,Fries:2003kq,Greco:2003xt,Hwa:2004ng,Fries:2008hs,Oh:2009zj,Fischer:2016zzs,Plumari:2017ntm,Zhao:2018jlw,He:2019vgs,He:2019tik,Cao:2019iqs,Minissale:2020bif,Andronic:2021erx,Zhao:2021vmu,Lonnblad:2023stc,Zhao:2024ecc,Ortiz:2026qmf,Vertesi:2026eul}.
In heavy-ion collisions, quark coalescence hadronization stands out as a widely used mechanism, which also accounts for the observed scaling behavior of elliptic flow
~\cite{Fries:2003kq,Molnar:2003ff,Greco:2003mm,Greco:2003vf,STAR:2004jwm,STAR:2005npq,PHENIX:2006dpn,PHENIX:2012swz}.
In small systems, by contrast, the enhancement has been attributed either to vacuum-like mechanisms acting on the color structure of the hard event itself, such as color reconnection and rope hadronization~\cite{Christiansen:2015yqa,Bierlich:2014xba}, or to the formation of a small QGP droplet in which hadrons are assembled by coalescence~\cite{Nagle:2018nvi}.
Whether the mechanisms behind this enhancement are the same across different collision systems and different flavor sectors remains an open question.

For the light-flavor sector, jet requirements have been used to separate particles produced in hard processes from those originating in the underlying event. The ALICE Collaboration measured $\Lambda$ and $\rm K_S^0$ production in jets~\cite{ALICE:2021vxl} and found that the in-jet $\Lambda/K^0_S$ ratio is lower than the inclusive one and does not exhibit the characteristic baryon enhancement at intermediate $p_T$, indicating that the light-flavor enhancement originates mainly from the underlying event. Supporting evidence was reported by the STAR Collaboration through measurements of the in-jet $p/\pi$ ratio at $\sqrt{s_{NN}}=200$~GeV~\cite{Dale-Gau:2023ree,Dale-Gau:2025nyw,STAR:2026jetratio}.

The heavy-flavor sector offers a particularly clean testing ground. The heavy-quark mass is much larger than the non-perturbative QCD scale, so heavy quarks are produced predominantly in initial hard scatterings and their numbers are conserved throughout the subsequent evolution. Nevertheless, the charmed baryon-to-meson ratio $\Lambda_c^+/D^0$ exhibits a significant enhancement at intermediate $p_T$ in p+p, p+Pb and Pb+Pb collisions~\cite{ALICE:2020wla,ALICE:2021bib,STAR:2019ank,CMS:2024tjo,CMS:2024Lamc}. Its multiplicity dependence in p+p collisions admits the same two competing classes of interpretations: vacuum-like mechanisms acting on the color structure of the hard event~\cite{Christiansen:2015yqa,Bierlich:2014xba}, and medium-like scenarios in which charmed baryons are assembled by coalescence in a small QGP droplet~\cite{Nagle:2018nvi}. Inclusive measurements alone cannot arbitrate between these two pictures, since both can describe the measured ratios.

Experimentally, the ALICE Collaboration has measured the parallel jet momentum fraction $z^{ch}_{||}$ distributions of charmed hadrons in p+p collisions~\cite{ALICE:2023jgm}; different PYTHIA8 tunes predict conflicting trends for the $\Lambda_c^+/D^0$ ratio, while the data favor the softer baryon fragmentation predicted by models with color correlations beyond the leading-color approximation. The ALICE Collaboration has also reported the radial distributions of $D^0$ mesons and $\Lambda_c^+$ baryons in p+p jets, suggesting that $\Lambda_c^+$ baryons are produced on average closer to the jet axis than $D^0$ mesons, a trend that the available model comparisons in that reference did not reproduce~\cite{Vertesi:2021brz}. More recently, the sensitivity of charm-tagged observables to the modelling of color reconnection has been examined with hadron-anchored energy correlators, which reveal a pronounced localized suppression of the energy flow around the $\Lambda_c^+$ relative to the $D^0$ in PYTHIA8, together with a qualitatively different color-reconnection dependence of the conventional jet-axis radial profile~\cite{Vertesi:2026wld}. Complementarily, the first measurement of azimuthal correlations between $\Lambda_c^+$ baryons and charged particles in p+p collisions hints at a larger yield of low-momentum particles associated with $\Lambda_c^+$ triggers than with $D$-meson triggers, a difference that current event generators likewise fail to reproduce~\cite{ALICE:2024lon}. Jets, as collimated sprays of hadrons originating from initial hard scatterings, also serve as penetrating probes of the quark-gluon plasma (QGP); heavy-flavor-tagged jets thus provide differential insights into the in-medium interactions of heavy quarks and into the hadronization mechanism itself.

In this context the jet requirement plays a twofold role. On the one hand, it removes the bulk (underlying-event) contribution; in the light-flavor sector, in-jet observables are therefore largely insensitive to a possible small-system medium, as supported empirically by the fact that the in-jet $\Lambda/K^0_S$ ratio in p+p is well described by PYTHIA8 without any medium~\cite{ALICE:2021vxl}. Charmed hadrons have no bulk component to remove, since charm quarks are produced in initial hard scatterings, but any medium-induced modification of charm hadronization would involve soft thermal partners uncorrelated with the jet and would be largely filtered out by the same requirement. On the other hand, the two classes of mechanisms are expected to imprint different spatial structures on the in-jet enhancement: the products of string-like mechanisms remain correlated with the jet, since they act on the color flow of the hard partons, whereas the products of coalescence with thermal partons carry no such correlation. The in-jet $\Lambda_c^+/D^0$ ratio and its radial profile therefore provide a handle on the origin of the enhancement that inclusive observables cannot provide. We note that this insensitivity argument rests on light-flavor observables; a dilute or short-lived small-system medium could still modify charm hadronization while leaving the light-flavor in-jet ratios unaffected. If future multiplicity-dependent in-jet measurements reveal a component of the $\Lambda_c^+/D^0$ enhancement that is not correlated with the jet axis, the p+p framework can be extended to include coalescence hadronization, as discussed in Sec.~VI.

In this work, we investigate the in-jet $\Lambda_c^+/D^0$ ratio in both p+p and Pb+Pb collisions as a discriminator of hadronization mechanisms. The motivation is that the inclusive $\Lambda_c^+/D^0$ ratio cannot distinguish color reconnection from in-medium coalescence---both mechanisms enhance it---whereas the jet-constrained observables considered here behave oppositely under the two scenarios. For the p+p baseline, charmed-hadron events are simulated with PYTHIA8~\cite{Bierlich:2022pfr} using two representative setups, the 4C tune and the CR-BLC Mode 2 tune~\cite{Christiansen:2015yqa}, the latter adopted as a representative beyond-leading-color hadronization scenario, and the light-flavor $\Lambda/K^0_S$ ratio serves as a validated benchmark. For Pb+Pb collisions we employ the updated SHELL model~\cite{Wang:2019xey,Wang:2020ukj,Xu:2025ivv} with a modified Langevin transport approach, incorporating both radiative and collisional energy loss, to describe the partonic evolution in the QGP medium, while a coalescence plus fragmentation hybrid approach~\cite{Xu:2025ivv} describes the in-medium hadronization of charm quarks. While the in-jet $z^{ch}_{||}$ of $\Lambda_c^+$ has been measured in p+p collisions~\cite{ALICE:2023jgm} and that of heavy-flavor mesons has been computed for Pb+Pb collisions~\cite{Li:2022hfmz}, to our knowledge, no calculation exists so far for jet-tagged charmed-hadron ratios in heavy-ion collisions that simultaneously includes medium-modified parton transport and in-medium hybrid hadronization. We provide predictions of the inclusive and in-jet $\Lambda_c^+/D^0$ ratios as functions of transverse momentum, the parallel jet momentum fraction $z^{ch}_{||}$ distributions of $\Lambda_c^+$ and $D^0$ decomposed into coalescence and fragmentation contributions together with the corresponding PbPb/pp ratios in the 0--10\% and 30--50\% centrality classes, and the $\Lambda_c^+/D^0$ ratio as a function of the distance from the jet axis; the $z^{ch}_{||}$ distributions obtained at $\sqrt{s}=5.02$~TeV are compared with the ALICE measurement at $\sqrt{s}=13$~TeV as a reference for their shape, since no measurement at 5.02~TeV is available. Sections~II and III describe the p+p baseline and the theoretical framework for Pb+Pb collisions, respectively. Section~IV studies the in-jet $\Lambda_c^+/D^0$ ratio in Pb+Pb and its radial profile, while Sec.~V presents the Pb+Pb $z^{ch}_{||}$ distributions and their radial profile, obtained in the kinematic window that matches the existing ALICE measurement. Section~VI contains a discussion of experimental caveats and possible extensions, followed by a summary.

\section{In-jet particle ratios in p+p collisions}
\begin{figure}[tbp]
\centering
\includegraphics[width=1.0\linewidth]{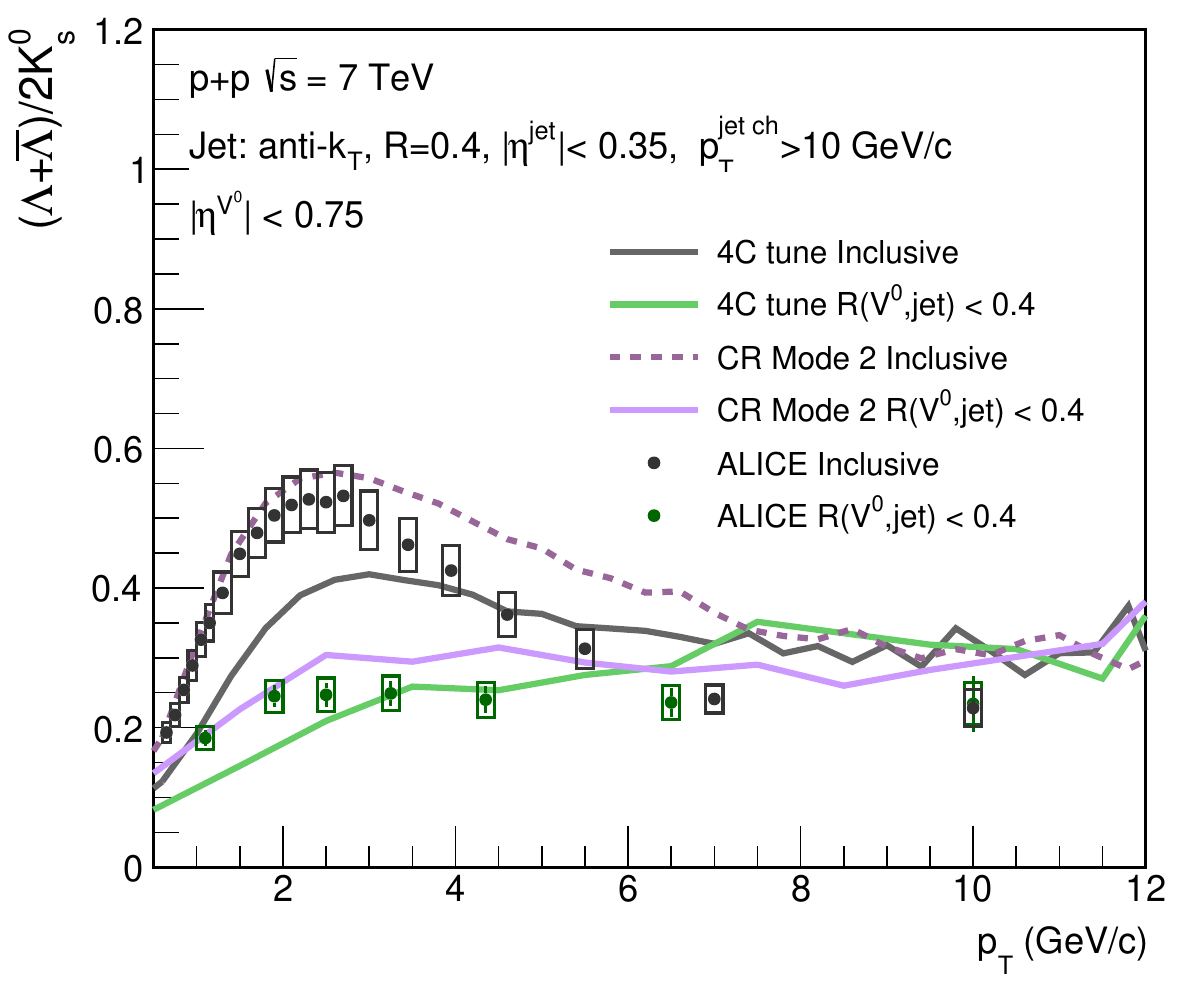}
\caption{ The $(\Lambda+\bar{\Lambda})/2K^0_S$ ratio in $\rm p+p$ collisions at $\sqrt{s}=7$ TeV as a function of $V^0$ particle $p_T$ calculated by PYTHIA8 4C and CR-BLC Mode 2, compared with ALICE~\cite{ALICE:2021vxl} data.}
\label{fig:light_flavor_ratio}
\end{figure}

In p+p collisions, events are generated with PYTHIA8~\cite{Bierlich:2022pfr} using the NNPDF2.3 parton distribution functions~\cite{Ball:2012cx}, with the SoftQCD processes and multiparton interactions switched on for both the light-flavor and heavy-flavor samples. Two hadronization setups are employed: the 4C tune~\cite{Corke:2010yf} and the CR-BLC Mode 2 tune~\cite{Christiansen:2015yqa} are both used in the light-flavor benchmark of Fig.~\ref{fig:light_flavor_ratio}, whereas all heavy-flavor results are obtained with CR-BLC Mode 2, adopted as a representative beyond-leading-color hadronization scenario. The 4C tune provides a standard Lund string fragmentation reference; CR-BLC Mode 2 implements color reconnection beyond the leading-color approximation, allowing the formation of baryonic configurations (string junctions), and improves the description of the baryon-to-meson ratios in p+p collisions~\cite{Bierlich:2015rha,ALICE:2020wla}.

The fragmentation-only baseline shown together with the heavy-flavor results is constructed from PYTHIA8 parton-level events, with charm quarks fragmented according to the Peterson function~\cite{Peterson:1982ak} (Sec.~III) and light quarks and gluons hadronized with PYTHIA8 string fragmentation, and serves as the reference for isolating the color-reconnection effects.

Charged-particle jets are reconstructed with the anti-$k_T$ algorithm ($R=0.4$) in FastJet~3~\cite{Cacciari:2011ma}. For the heavy-flavor observables, a jet is retained if it is associated with at least one prompt charmed hadron, and a charmed hadron $h$ is counted as in-jet when its distance from the jet axis satisfies $R(\mathrm{jet},h)<0.4$. For the light-flavor benchmark, no hadron-content requirement is imposed on the reconstructed jets, and a $V^0$ is counted as in-jet when $R(\mathrm{jet},V^0)<0.4$.

\begin{figure}[tbp]
\centering
\includegraphics[width=1.0\linewidth]{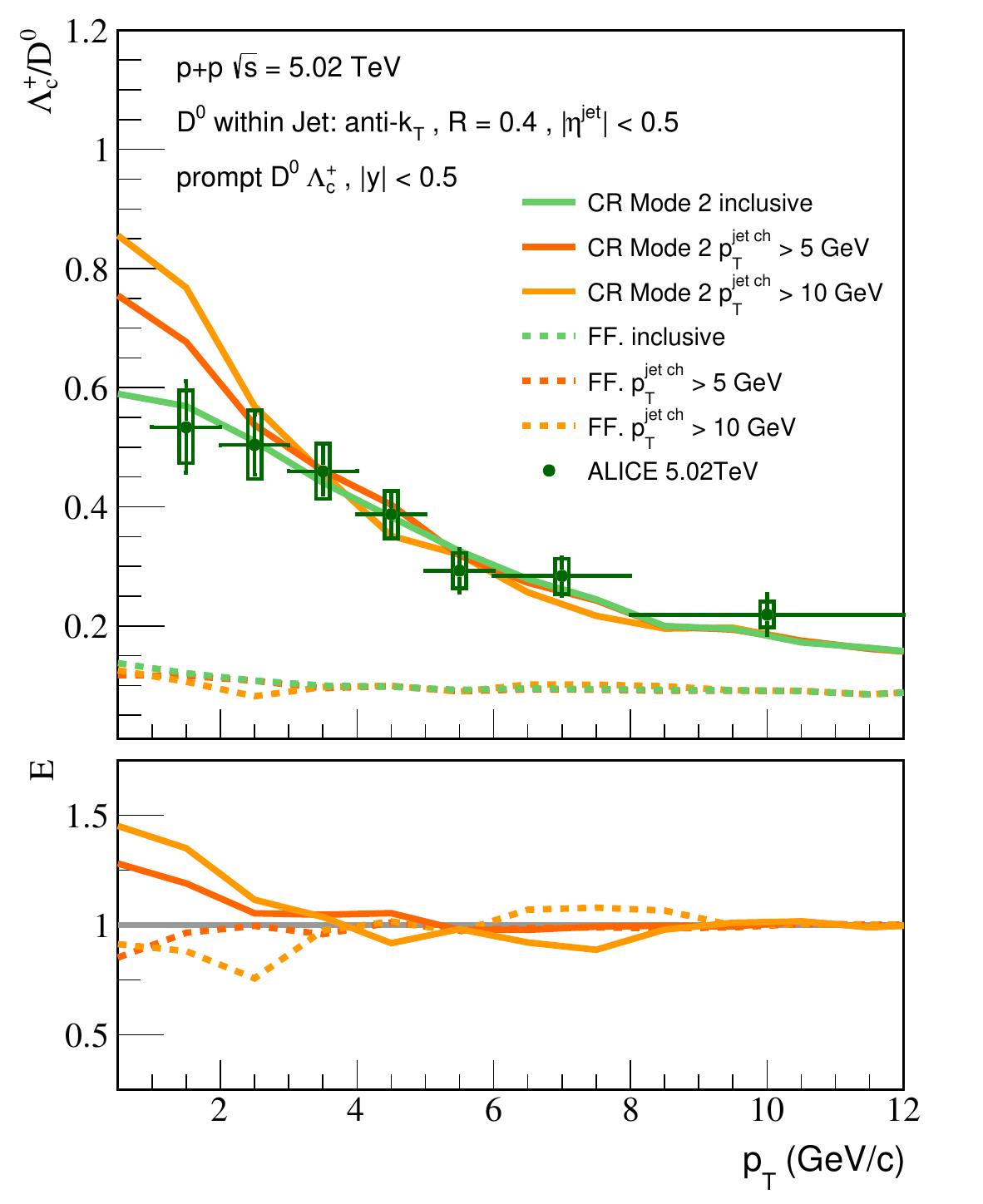}
\caption{ The inclusive and in-jet $\Lambda_c^+/D^0$ ratios as a function of the heavy-hadron transverse momentum $p_T$ for two jet $p_T$ selections, $p_T^{\rm ch\,jet}>5$ and $10$~GeV/$c$ in $\mathrm{p+p}$ collisions at $\sqrt{s}=5.02$~TeV. The results are calculated with PYTHIA8 CR-BLC Mode 2 and compared to ALICE data~\cite{ALICE:2020wla}. Jets are reconstructed with the anti-$k_T$ algorithm with $R=0.4$ and $|\eta_{\rm jet}|<0.5$.}
\label{fig:ratio-5020-pp-CR2-LcD0}
\end{figure}

Two complementary kinematic windows are employed throughout this work. The particle-ratio observables, including the radial profiles, use jets with $p_T^{\rm ch\,jet}>7$~GeV/$c$ and no upper limit, together with charmed hadrons in $0.5<p_T^h<15$~GeV/$c$; this low-$p_T$ window covers the region where the charmed-baryon enhancement resides and is used to study the sign of the in-jet ratio modification (Fig.~\ref{fig:ratio-5020-pp-CR2-LcD0}) and its radial profile around the jet axis. The $z^{ch}_{||}$ observables use a harder window, $7<p_T^{\rm ch\,jet}<15$~GeV/$c$ and $3<p_T^h<15$~GeV/$c$, which matches the kinematics of the ALICE measurement~\cite{ALICE:2023jgm} and allows the coalescence and fragmentation contributions to be examined separately and confronted directly with data.

As a benchmark, Fig.~\ref{fig:light_flavor_ratio} shows the light-flavor $(\Lambda+\bar{\Lambda})/2K^0_S$ ratio as a function of hadron $p_T$, computed at $\sqrt{s}=7$~TeV to match the ALICE measurement~\cite{ALICE:2021vxl}. Both tunes yield an in-jet ratio below the inclusive one at intermediate $p_T$, in agreement with the data. In the low-to-intermediate $p_T$ region, CR Mode 2 gives a significantly higher ratio than the 4C tune in both the inclusive and in-jet samples, indicating that the light-flavor baryon-to-meson enhancement originates mainly from soft processes outside the jet, consistent with the experimental interpretation~\cite{ALICE:2021vxl}.

\noindent\textbf{\textit{In-jet charmed baryon-to-meson ratio.}}---Figure~\ref{fig:ratio-5020-pp-CR2-LcD0} shows the inclusive and in-jet $\Lambda_c^+/D^0$ ratios as functions of the hadron $p_T$ for two jet-$p_T$ selections. The ALICE data shown for comparison correspond to the inclusive measurement, without any jet selection~\cite{ALICE:2020wla}; our inclusive ratio is consistent with them, supporting the CR-BLC Mode 2 description of charm hadronization in p+p collisions. In the heavy-flavor sector, the effect of the jet requirement is opposite to the light-flavor case: at low-to-intermediate $p_T$, the in-jet ratio lies above the inclusive one. Quantifying the effect by the enhancement factor $E\equiv R_{\rm jet}/R_{\rm incl}$, we find $E-1>0$ at low-to-intermediate $p_T$, with the enhancement growing for increasing jet-$p_T$ selection, while $E$ approaches unity at high $p_T$. 

The enhancement arises because the color-reconnection excess is built from the color flow of the hard partons: its products remain correlated with the jet and survive the jet requirement, whereas hadrons uncorrelated with the jet are discarded, driving the in-jet ratio above the inclusive one. The fragmentation-only baseline shows no jet-induced change of the ratio ($E_{\rm FF}\simeq1$). In what follows, we refer to the excess of the ratio over the fragmentation-only baseline as the color-reconnection excess; the in-jet enhancement is specific to this excess.

\begin{figure}[tbp]
\centering
\includegraphics[width=0.9\linewidth]{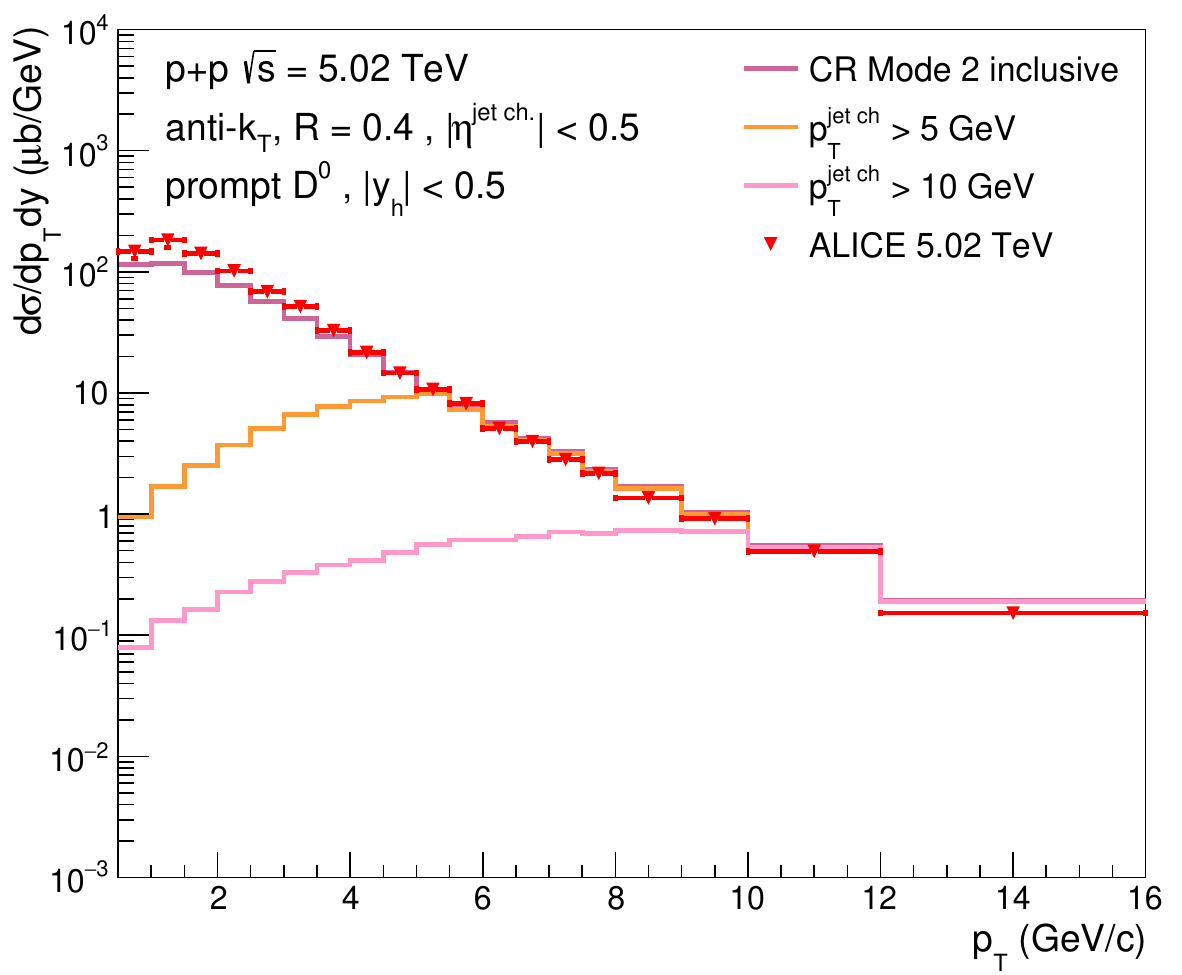}
\includegraphics[width=0.9\linewidth]{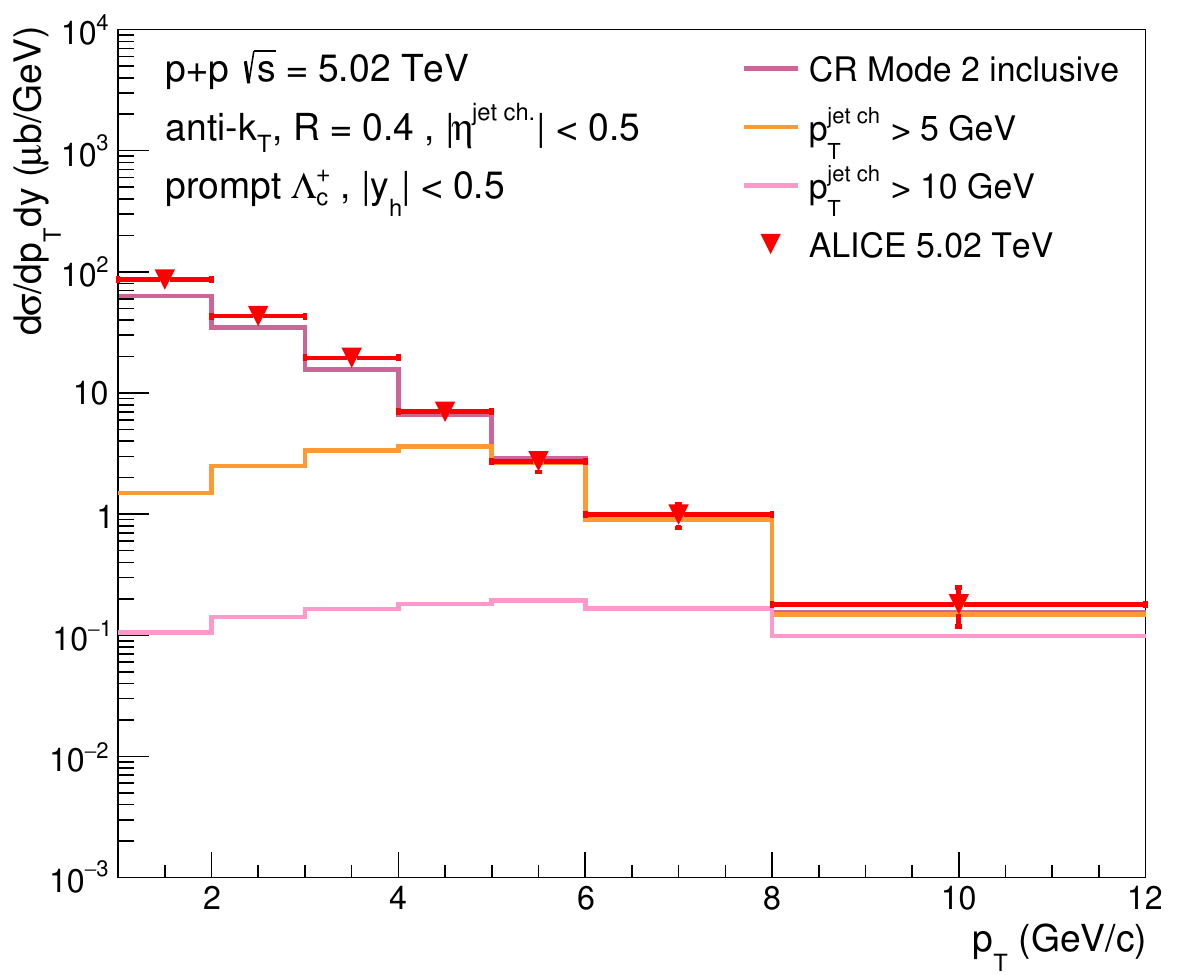}
\caption{Inclusive and in-jet $p_T$-differential production cross sections for prompt $D^0$ (upper panel) and prompt $\Lambda_c^+$ (lower panel) in $\mathrm{p+p}$ collisions at $\sqrt{s}=5.02$~TeV. The in-jet results are shown for two jet $p_T$ thresholds, $p_T^{\rm jet}>5$~GeV/$c$ and $p_T^{\rm jet}>10$~GeV/$c$. The results are calculated with PYTHIA8 CR-BLC Mode 2 and compared to ALICE data~\cite{ALICE:2021mgk,ALICE:2020wla}.}
\label{fig:pp_jetspec}
\end{figure}

To illustrate the effect of the jet requirement on the spectra of the two different final-state hadrons, we show in Fig.~\ref{fig:pp_jetspec} the inclusive and in-jet $p_T$-differential cross sections of $D^0$ and $\Lambda_c^+$ for two jet-$p_T$ selections, simulated with PYTHIA8 CR-BLC Mode 2. At low $p_T$, a larger fraction of $D^0$ mesons than of $\Lambda_c^+$ baryons is rejected by the jet requirement. This species-dependent rejection is what makes the in-jet $\Lambda_c^+/D^0$ ratio exceed the inclusive one at low-to-intermediate $p_T$; in the following we examine how it depends on the distance from the jet axis.

\begin{figure}[tbp]
\centering
\includegraphics[width=0.9\linewidth]{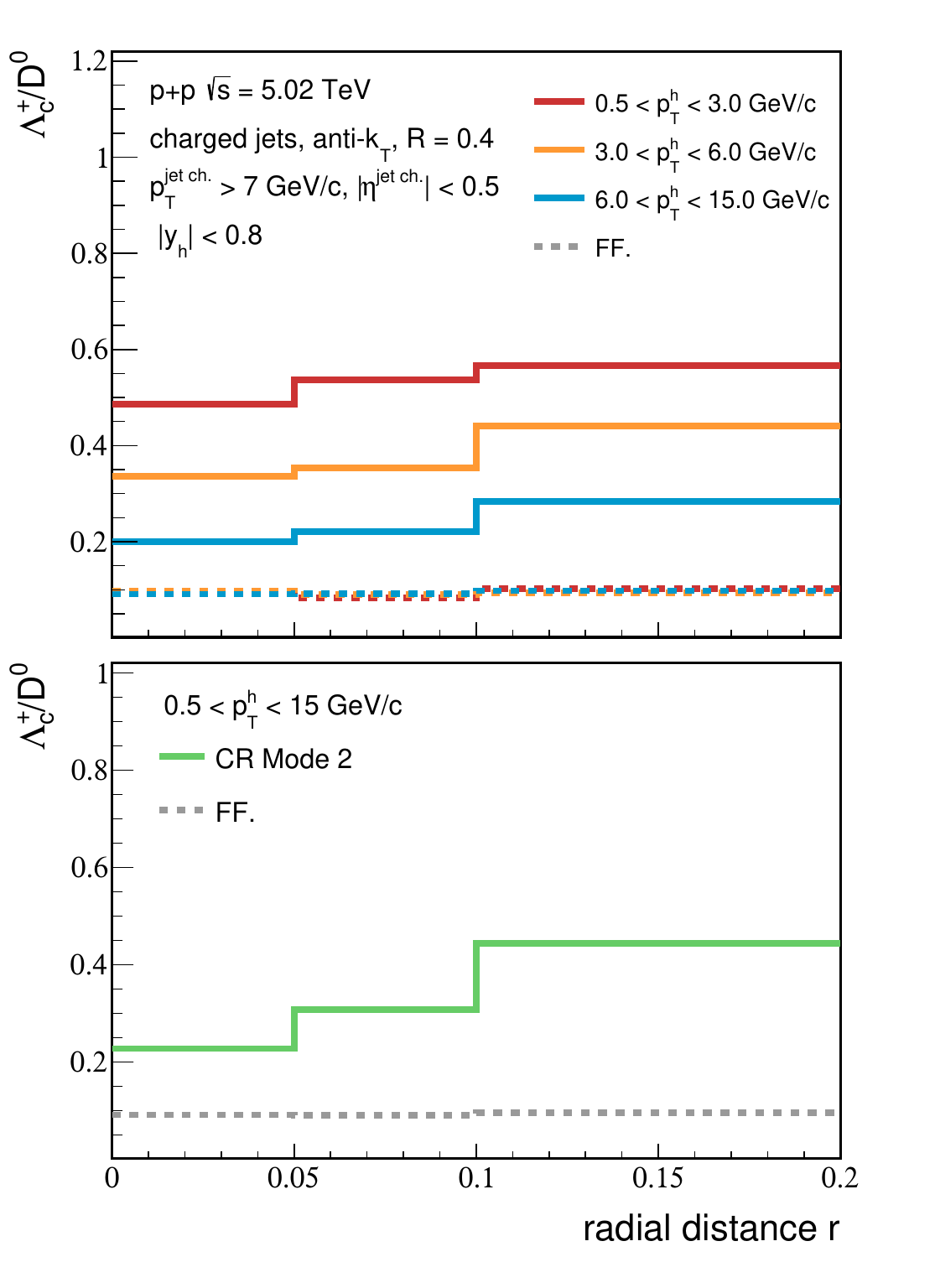}
\caption{ $\Lambda_c^+/D^0$ ratios as a function of the distance from the jet axis in $\mathrm{p+p}$ collisions at $\sqrt{s}=5.02$~TeV simulated by PYTHIA8 CR-BLC Mode 2. Jets are reconstructed with the anti-$k_T$ algorithm with $R=0.4$ and are selected with $p_T^{\text{ch jet}} > 7$~GeV/$c$ and $|\eta_{\text{jet}}| < 0.5$, and the charmed hadrons are required to be prompt with $|y_h| < 0.8$.
The ratios are presented for three charmed-hadron $p_T$ intervals, $0.5$--$3.0$, $3.0$--$6.0$ and $6.0$--$15.0$~GeV/$c$. Dashed lines correspond to the fragmentation-only baseline. The distributions are shown up to $r=0.2$; at larger distances from the jet axis the present samples do not provide sufficient statistical precision for a stable ratio. }
\label{fig:R_jet_V0-pp}
\end{figure}

\noindent\textbf{\textit{Radial distribution around the jet axis.}}---The radial distribution of hadrons around the jet axis carries complementary information on how the production mechanism reshapes the spatial structure of the low-$p_T$ enhancement in the transverse plane. Figure~\ref{fig:R_jet_V0-pp} shows the $\Lambda_c^+/D^0$ ratio as a function of the distance between the charmed hadron and the jet axis, $R(\mathrm{jet},h)=\sqrt{(\Delta\eta)^2+(\Delta\phi)^2}$, for three charmed-hadron $p_T$ intervals in p+p collisions simulated with the CR-BLC Mode 2 tune. 
Within the jet, the ratio exhibits the same hierarchy seen in the Figure~\ref{fig:ratio-5020-pp-CR2-LcD0}: highest in the low-$p_T$ interval, $0.5<p_T^h<3.0$~GeV/$c$, and lowest in the high-$p_T$ interval, $6.0<p_T^h<15.0$~GeV/$c$. In all three intervals, it increases with distance from the jet axis,
while the fragmentation-only baseline is essentially independent of the distance from the jet axis. This indicates that the color-reconnection excess remains correlated with the jet but is displaced towards larger distances from the axis, 
so that the additional baryon yield is not concentrated in the jet core. 
This pattern reflects how the jet requirement filters the spatial structure of the low-$p_T$ enhancement: the color-reconnection products can remain associated with a hard jet only when they are emitted at larger angles from its axis, 
whereas fragmentation products here are not subject to the kind of spatial filtering that differentiates $D^0$ and $\Lambda^+_c$ and therefore they do not exhibit the special radial structure induced by color reconnection discussed above.

The ALICE Collaboration has reported that $\Lambda_c^+$ baryons are, on average, closer to the jet axis than $D^0$ mesons, a trend that the model calculations discussed in that measurement did not reproduce~\cite{Vertesi:2021brz}. 
The CR-BLC Mode 2 prediction points in the opposite direction---the $\Lambda_c^+/D^0$ ratio increases with the distance from the jet axis---but a direct confrontation is not yet possible: the two analyses differ in kinematic windows and jet definitions, and establishing whether the two descriptions can be reconciled requires a dedicated comparison within a common setup. Complementary information on the local hadronization dynamics can be obtained from hadron-anchored energy correlators, for which the conventional jet-axis radial profile has been shown to exhibit a qualitatively different color-reconnection dependence~\cite{Vertesi:2026wld}.
These p+p results will serve as the baseline for the study of medium modifications in Pb+Pb collisions presented in Secs.~IV and V.

\noindent\textbf{\textit{Parallel jet momentum fraction.}}---Throughout this work, the parallel jet momentum fraction of a charmed hadron $h$ with respect to its associated charged-particle jet is defined as
\begin{eqnarray}
z^{ch}_{||}=\dfrac{\vec{p}_{\mathrm{jet}}\cdot\vec{p}_{h}}{\vec{p}_{\mathrm{jet}}\cdot\vec{p}_{\mathrm{jet}}}=\dfrac{|\vec{p}_{h}|}{|\vec{p}_{\mathrm{jet}}|}\cos\theta,
\label{eqn:zpara}
\end{eqnarray}
where $\theta$ is the angle between the charmed-hadron momentum and the jet axis, and the superscript ``ch'' indicates that the jets are reconstructed from charged particles. The corresponding distributions are normalized per jet, $(1/N_{\mathrm{jet}})\,dN/dz^{ch}_{||}$.

\begin{figure}[tbp]
\centering
\includegraphics[width=1.0\linewidth]{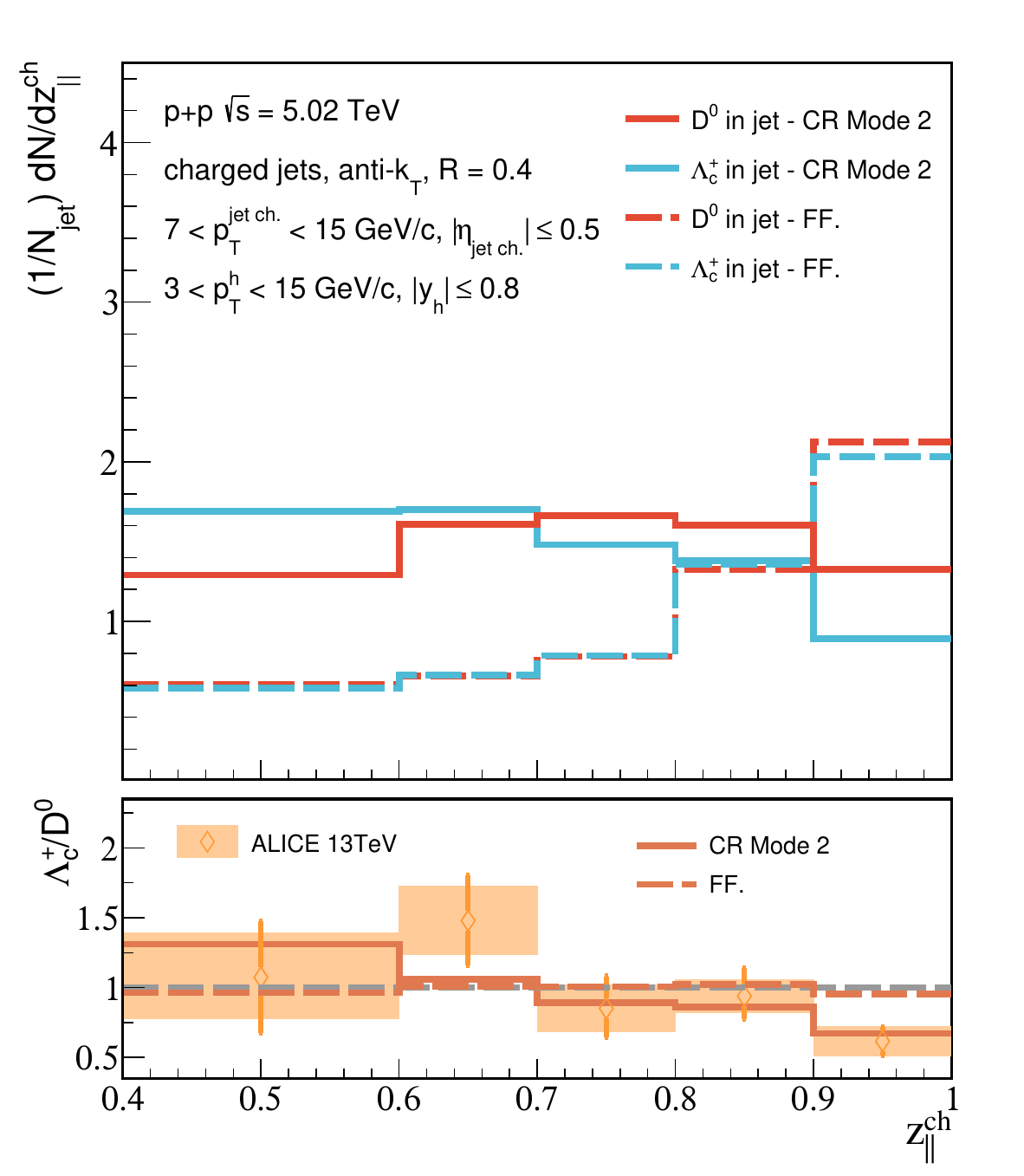}
\caption{ $\Lambda_{c}^{+}$ and $D^{0}$ parallel jet momentum fraction distributions in $\mathrm{p+p}$ collisions at $\sqrt{s}=5.02$~TeV, calculated with PYTHIA8 CR-BLC Mode 2 (solid lines) and with the fragmentation-only baseline (dashed lines) in the kinematic range of the ALICE measurement~\cite{ALICE:2023jgm}: jets are selected with $|\eta_{\rm jet}|<0.5$ and $7<p_T^{\rm ch\,jet}<15$~GeV/$c$, and prompt charmed hadrons with $3<p_T^h<15$~GeV/$c$ and $|y_h|<0.8$. The bottom panel shows the corresponding $\Lambda_c^+/D^0$ ratio together with the ALICE measurement at $\sqrt{s}=13$~TeV~\cite{ALICE:2023jgm}, which is used here as a reference for the shape since no measurement at 5.02~TeV is available.}
\label{fig:zpara-pp-exp}
\end{figure}

Figure~\ref{fig:zpara-pp-exp} shows the $\Lambda_c^+$ and $D^0$ $z^{ch}_{||}$ distributions in p+p collisions at $\sqrt{s}=5.02$~TeV in the kinematic range of the ALICE measurement~\cite{ALICE:2023jgm}, for CR-BLC Mode 2 (solid) and the fragmentation-only baseline (dashed), with their ratio compared to the ALICE 13~TeV data in the bottom panel. Relative to the baseline, both spectra are shifted toward small $z^{ch}_{||}$, with a more pronounced shift for $\Lambda_c^+$, so that the ratio decreases with $z^{ch}_{||}$---opposite to the baseline ratio, which shows only a marginal upward trend consistent with the PYTHIA8 Monash results reported in Ref.~\cite{ALICE:2023jgm}---and indicating that $\Lambda_c^+$ production is softer and/or less aligned with the jet axis under color reconnection. The decreasing CR-BLC prediction is compatible with the ALICE data within the sizable experimental uncertainties, which do not yet discriminate between the two scenarios; the 13~TeV data serve as a shape reference only, since no measurement at 5.02~TeV is available.

We note for later discussion that a decreasing $\Lambda_c^+/D^0$ ratio with $z^{ch}_{||}$ is a generic feature of recombination-type hadronization(color reconnection and coalescence) rather than a fingerprint of a specific mechanism: as shown in Sec.~V, the coalescence mechanism in Pb+Pb produces the same trend. The slope of the ratio therefore identifies \emph{whether} non-fragmentation physics is at work, while the identity of the mechanism is resolved by combining the $z^{ch}_{||}$ spectrum with the sign of the in-jet $p_T$ ratio and with the radial profile.

\section{Framework for Pb+Pb collisions}
The p+p baseline of Sec.~II showed that color reconnection modifies all three in-jet observables: the $\Lambda_c^+/D^0$ ratio, its radial profile, and the $z^{ch}_{||}$ spectra. In Pb+Pb collisions, the QGP medium enhances coalescence and introduces parton energy loss, both of which can alter the momentum balance inside jets. In this section we present the simulation framework for Pb+Pb collisions; its application to the in-jet ratio and its radial profile and to the $z^{ch}_{||}$ observables follows in Secs.~IV and V.

\subsection{Initial conditions and in-medium parton transport}
In Pb+Pb collisions,
the parton-level events per nucleon-nucleon collision are simulated by PYTHIA8 with the same setup as described in Sec.~II. The initial spatial positions of the partons are taken from the binary collision vertices provided by the Glauber Monte Carlo model~\cite{Miller:2007ri}.

The SHELL model is employed to simulate the in-medium evolution and energy loss of partons. For heavy quarks propagating through the hot and dense medium, a modified Langevin equation is adopted, which simultaneously incorporates collisional and radiative energy loss.
\begin{eqnarray}
	dx_j & = & \dfrac{p_j}{E}dt ,\label{eqn:dxj} \\
	dp_j & = & -\Gamma p_jdt+\sqrt{\kappa dt}\rho_j -\vec{p}_g \label{eqn:dpj}
\end{eqnarray}
$j=1,2,3$ labels the three spatial directions. $\Gamma$ is the drag coefficient which constrains the strength of collisional interaction. The $\sqrt{\kappa dt}\rho_j$ term describes the random kicks from medium partons. $\rho_j$ is a Gaussian random variable with zero mean and unit variance. The diffusion coefficient $\kappa$ satisfies the fluctuation-dissipation relation with $\Gamma$~\cite{Kubo:1966fyg}
\begin{eqnarray}
	\kappa=2\Gamma ET=\dfrac{2T^2}{D_s} \label{eqn:fluc-diss}
\end{eqnarray}
The spatial diffusion coefficient is set to a constant, $2\pi T D_s=2.5$, taken from the (2+1)-flavor lattice QCD calculation~\cite{Altenkort:2023oms}.
The collisional energy loss of light quarks is treated using perturbative QCD in the Hard Thermal Loop approximation~\cite{Neufeld:2010xi}.
\begin{eqnarray}
	\dfrac{dE^{coll}}{dt}=\dfrac{\alpha_sC_s\mu^2_D}{2}\ln\dfrac{\sqrt{ET}}{\mu_D} \label{eqn:HTL}
\end{eqnarray}
where $C_s$ is the Casimir factor in the fundamental or adjoint representation and $\mu_D$ is the Debye screening mass.

For both heavy and light quarks, the radiative contribution is accounted for through a discrete momentum recoil term $\vec{p}_g$ arising from medium-induced gluon radiation.
The Higher-Twist approach is adopted in the SHELL model~\cite{Guo:2000nz,Zhang:2003wk,Zhang:2004qm,Majumder:2009ge} where the differential radiation spectrum for the emitted gluon reads
\begin{eqnarray}
    \dfrac{dN}{dxdk^2_{\perp}dt}=\dfrac{2\alpha_sC_sP(x)\hat{q}}{\pi k^4_{\perp}}\sin^2(\dfrac{t-t_i}{2\tau_f})(\dfrac{k^2_{\perp}}{k^2_{\perp}+x^2m^2})^4 \label{eqn:gluon radiation spectra}
\end{eqnarray}
Here $x$ and $k_{\perp}$ denote the energy fraction and transverse momentum of the gluon radiated from the mother parton, respectively. $P(x)$ is the splitting function in vacuum~\cite{Deng:2009ncl}. The formation time of the radiated gluon is given by $\tau_f=2Ex(1-x)/(k^2_{\perp}+x^2m^2)$. The jet transport parameter $\hat{q}$ in the QGP is assumed to be proportional to $(T/T_0)^3$ in the hot QCD medium~\cite{Chen:2010te}, namely $\hat{q}(\tau,r)=\hat{q}_0(T/T_0)^3 (p_{\mu}u^{\mu})/p^0$, with $T_0=165$~MeV. For LHC energies, we adopt $\hat{q}_0 = 1.5{~\rm GeV^2/fm}$, as suggested by the $\chi^2$ analysis of single-hadron and dihadron nuclear modification~\cite{Xie:2019oxg}.
In the present work, the space-time evolution of the QGP medium is obtained from the (3+1)-dimensional viscous hydrodynamic model CLVisc~\cite{Pang:2018zzo,Wu:2021fjf}.

\subsection{Hadronization}
When a local fluid cell cools down to the critical temperature $T_c=165$~MeV, hadronization is triggered.
For charm quarks, we adopt a coalescence plus fragmentation hybrid model to describe their hadronization~\cite{Xu:2025ivv}.
The momentum spectrum of hadrons produced via coalescence on the hadronization hypersurface is given by
\begin{eqnarray}
\label{coalescence}
{dN\over d^3{\bf P}}&=&g \int \prod_{i=1}^n{d^3x_id^3p_i\over (2\pi)^3E_i} f_i({\bf x}_i,{\bf p}_i)\\
&\cdot& W({\bf x}_1,\cdots,{\bf x}_n, {\bf p}_1,\cdots, {\bf p}_n)\, \delta^{(3)}\left({\bf P}-\sum_{i=1}^n{\bf p}_i\right),\nonumber
\end{eqnarray}
where $g$ is the statistical factor for the spin and color degrees of freedom, $i$ is the index of the $i$-th constituent quark of the produced hadron. $f_i(x_i,p_i)$ denotes the phase-space distribution function of quark $i$.
Light quarks are taken to be in local thermal equilibrium with the QGP medium, and thus their phase-space distribution follows the Fermi-Dirac form $f_i(r_i,p_i) = g_i/(e^{u_\mu p^\mu_i/T} + 1)$, where the local fluid four-velocity $u_\mu(x_i)$ and temperature $T(x_i)$ are supplied by the hydrodynamic simulation~\cite{Pang:2018zzo,Wu:2021fjf}.
$W$ is the Wigner function of the charmed hadron, obtained from the harmonic oscillator wave function with properly chosen width parameters~\cite{Zhao:2025cnp}.
The total coalescence probability is obtained by summing the integrated spectra of all involved charm hadron species. A normalization constant $c$ is introduced to ensure that this total probability approaches unity as the charm-quark momentum tends to zero. This reflects the assumption that at very low momentum in the fluid rest frame, the charm quark is guaranteed to form a hadron via coalescence.

The charm quarks that are not selected through the total coalescence probability are assumed to hadronize through fragmentation, where the momentum fraction $z$ of the produced hadron follows the Peterson fragmentation function~\cite{Peterson:1982ak}
\begin{eqnarray}
	\mathscr{D}_{c\to H}(z) \propto \dfrac{1}{z[1-\dfrac{1}{z}-\dfrac{\epsilon}{1-z}]^2}
	\label{eqn:Peterson}
\end{eqnarray}
The Peterson parameter is taken as $\epsilon=0.01$ for $D^0$ and $\epsilon=0.02$ for $\Lambda_c^+$. The fragmentation fractions of charm quarks into different hadron species are extracted from experimental data in $\rm p+p$ collisions~\cite{Lisovyi:2015uqa}.

For light partons that have propagated through the medium and reached the hadronization hypersurface, we adopt a colorless hadronization procedure~\cite{Putschke:2019yrg}: quark-antiquark pairs are formed by minimizing their distance in the $(\eta,\phi)$ space. Gluons are then attached to the nearest color string. Once the color flow for all strings is constructed, partons are passed to PYTHIA8 to perform the hadronization.

\subsection{Jet reconstruction}
Charged-particle jets are reconstructed with the anti-$k_T$ algorithm ($R=0.4$) in FastJet~3~\cite{Cacciari:2011ma}, with the same heavy-flavor jet requirement as in p+p (Sec.~II): a jet is retained if it is associated with at least one prompt charmed hadron within $R(\mathrm{jet},h)<0.4$. Since the calculation is a full event-by-event simulation in which the medium evolution and the soft sector are modeled explicitly, the reconstructed jets contain only simulated final-state particles, and no background subtraction is applied. The kinematic selections for each observable are given in the corresponding figure captions.

\section{In-jet $\Lambda_c^+/D^0$ ratio and its radial profile in Pb+Pb collisions}
\begin{figure}[tbp]
\centering
\includegraphics[width=0.94\linewidth]{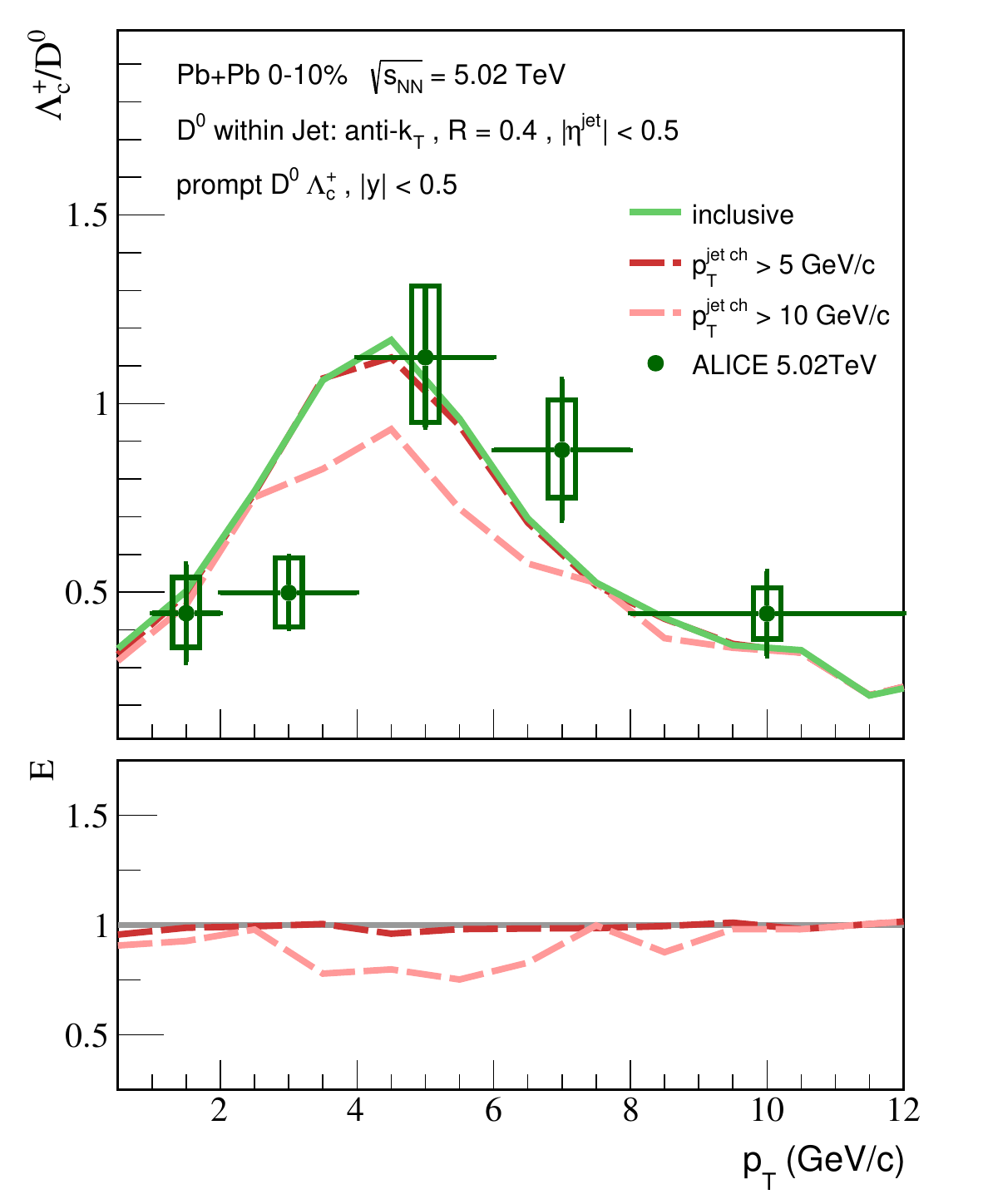}
\caption{ The inclusive and in-jet $\Lambda_c^+/D^0$ ratios as a function of the heavy-hadron transverse momentum $p_T$ for two jet $p_T$ selections, $p_T^{\rm ch\,jet}>5$ and $>10$~GeV/$c$, in 0-10\% central $\mathrm{Pb+Pb}$ collisions at $\sqrt{s_{NN}}=5.02$~TeV, compared to ALICE data~\cite{ALICE:2021bib}. Jets are reconstructed with the anti-$k_T$ algorithm with $R=0.4$, $|\eta_{\rm jet}|<0.5$, and the charmed hadrons are required to be prompt with $|y|<0.5$. The two thresholds illustrate the dependence on the jet selection; the reference selection adopted for the radial profile of Fig.~\ref{fig:R_jet_V0} is $p_T^{\rm ch\,jet}>7$~GeV/$c$.}
\label{fig:ratio-5020-0_10-LcD0}
\end{figure}

Following the p+p study of Sec.~II, we ask whether the enhancement of the baryon-to-meson ratio at intermediate $p_T$ arises from hard processes associated with jets. Figure~\ref{fig:ratio-5020-0_10-LcD0} shows the effect of the jet requirement on the $\Lambda_c^+/D^0$ ratio in 0-10\% central Pb+Pb collisions for two jet $p_T$ thresholds; the inclusive ratio is consistent with the ALICE measurement within the experimental uncertainties. Relative to the inclusive case, the in-jet ratio is suppressed at intermediate $p_T$: for $p_T^{\rm ch\,jet}>10$~GeV/$c$ the suppression amounts to 10--20\% in the $4<p_T^h<8$~GeV/$c$ region, i.e. $E<1$, whereas the softer $p_T^{\rm ch\,jet}>5$~GeV/$c$ selection reduces it only slightly. The reference selection used below for the radial observables, $p_T^{\rm ch\,jet}>7$~GeV/$c$ with no upper limit, lies between the two thresholds.

The in-jet modification therefore has the opposite sign in the two systems (Sec.~II). The origin of this sign flip lies in the different ways in which the jet requirement reshapes the ratio. In p+p, the fragmentation-only baseline develops no low-$p_T$ enhancement, in either the inclusive or the in-jet ratio ($E_{\rm FF}\simeq1$), whereas color reconnection does and the jet requirement raises the ratio further ($E-1>0$ at low-to-intermediate $p_T$, Sec.~II). In Pb+Pb, by contrast, the coalescence component is assembled with soft thermal partons whose momenta are uncorrelated with the jet direction, so that it is depleted in the jet-selected sample and the in-jet ratio falls back towards its fragmentation-only value. The dependence on the jet threshold in Fig.~\ref{fig:ratio-5020-0_10-LcD0} has a further, momentum-dependent origin: a harder jet requirement selects harder charm quarks, for which the coalescence probability is smaller, so that the depletion is weak for the soft $p_T^{\rm ch\,jet}>5$~GeV/$c$ selection and pronounced for the harder ones. The sign of the in-jet modification thus provides a discriminator of the hadronization scenario.

\begin{figure}[tbp]
\centering
\includegraphics[width=1.0\linewidth]{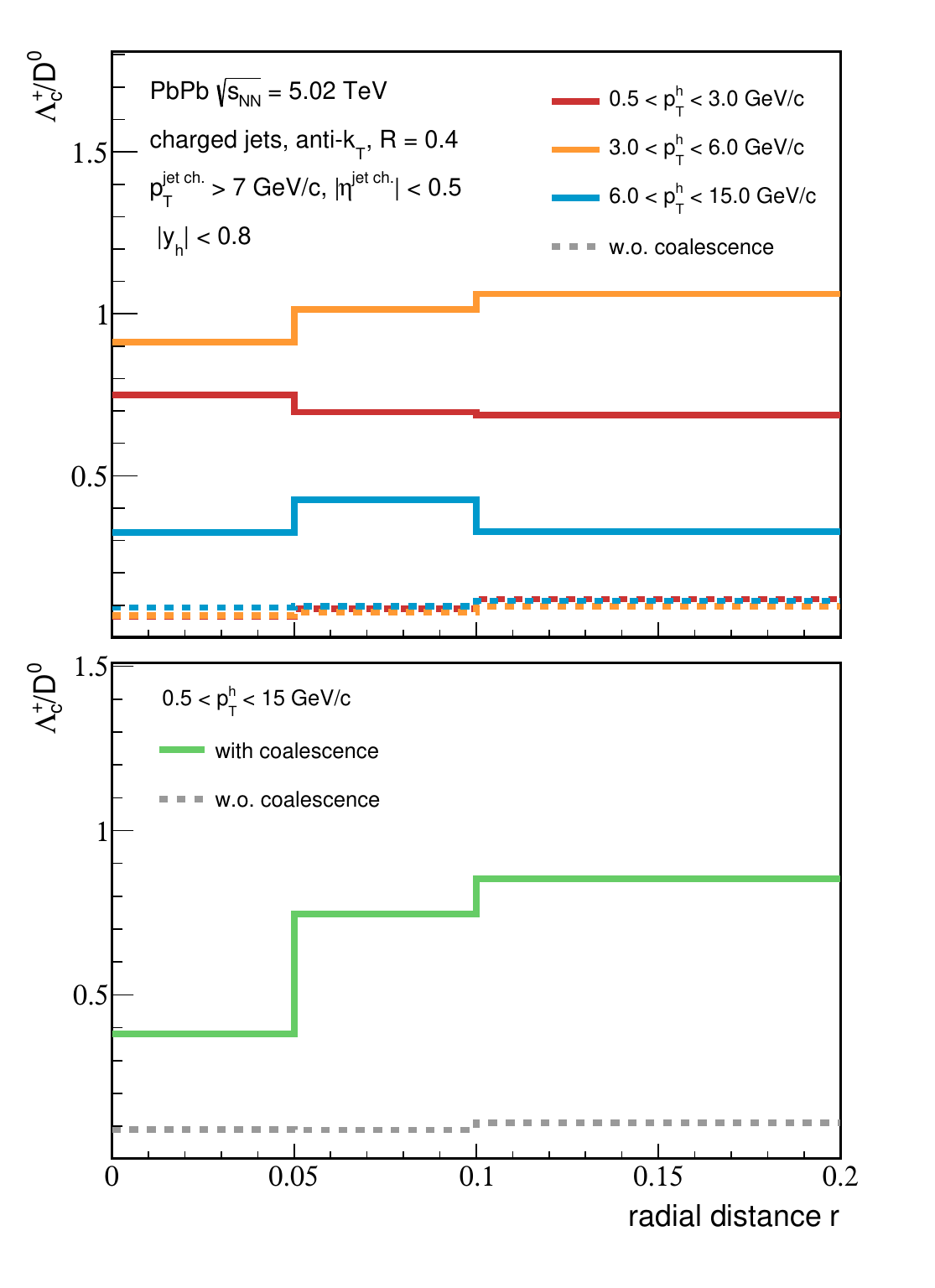}
\caption{
$\Lambda_c^+/D^0$ ratios as a function of the distance from the jet axis in 0-10\% central $\mathrm{Pb+Pb}$ collisions at $\sqrt{s_{NN}}=5.02$~TeV, reconstructed with the anti-$k_T$ algorithm with $R=0.4$. Jets are selected with $p_T^{\text{ch jet}} > 7$~GeV/$c$ and $|\eta_{\text{jet}}| < 0.5$.
The ratios are presented for three charmed-hadron $p_T$ intervals, $0.5$--$3.0$, $3.0$--$6.0$ and $6.0$--$15.0$~GeV/$c$, with $|y_h|<0.8$.
Dashed lines correspond to charmed hadrons produced when only fragmentation hadronization is enabled (coalescence turned off). The distributions are shown up to $r=0.2$; at larger distances from the jet axis the present samples do not provide sufficient statistical precision for a stable ratio. }
\label{fig:R_jet_V0}
\end{figure}

This interpretation can be tested directly in the spatial distribution of the in-jet ratio. Figure~\ref{fig:R_jet_V0} shows the $\Lambda_c^+/D^0$ ratio as a function of the distance from the jet axis in 0-10\% central Pb+Pb collisions, using the same jet selection as the p+p radial profile of Sec.~II ($p_T^{\rm ch\,jet}>7$~GeV/$c$, $|\eta_{\rm jet}|<0.5$). With fragmentation only (coalescence switched off), the ratios for the three charmed-hadron $p_T$ intervals are essentially independent of the distance from the jet axis. Once coalescence is included, the ratio increases with the distance from the axis over the range shown ($r\lesssim0.2$), most strongly for the intermediate interval $3<p_T^h<6$~GeV/$c$ and only weakly for the lowest and highest intervals.
After integrating over $0.5<p_T^h<15$~GeV/$c$, the total in-jet $\Lambda_c^+/D^0$ ratio shows a modest increase with distance from the jet axis. 
This behaviour is consistent with a coalescence component assembled with soft thermal partons that are uncorrelated with the jet direction: the additional yield is displaced towards larger distances from the axis rather than concentrated around it, and it is strongest where coalescence contributes most to the ratio.

A comparison with the p+p profile of Fig.~\ref{fig:R_jet_V0-pp} shows that in both collision systems the additional yield is displaced towards larger distances from the axis, but that the $p_T$ interval displaced farthest outward differs: the lowest-$p_T$ interval for color reconnection and the intermediate-$p_T$ interval for coalescence. The radial profile of the in-jet $\Lambda_c^+/D^0$ ratio thus provides information complementary to its $p_T$ and $z^{ch}_{||}$ dependences, and a $p_T$-differential measurement in both collision systems would directly test the mechanism-dependent spatial patterns predicted here.

\section{Parallel jet momentum fraction and its radial profile in Pb+Pb collisions}
As shown in Sec.~IV, for jets with $p_T^{\rm ch\,jet}>10$~GeV/$c$ the in-jet $\Lambda_c^+/D^0$ ratio is suppressed relative to the inclusive one, and the additional yield is displaced towards larger distances from the axis, most strongly at intermediate $p_T$, because the coalescence products are uncorrelated with the jet direction. We now turn to the longitudinal momentum-fraction dimension of the same question, in the jet window $7<p_T^{\rm ch\,jet}<15$~GeV/$c$ with $3<p_T^h<15$~GeV/$c$ and $R=0.4$, which matches the kinematics of the existing p+p $z^{ch}_{||}$ measurement, so that the corresponding Pb+Pb predictions can be confronted with data once a Pb+Pb measurement at the same energy becomes available.

\begin{figure}[tbp]
\centering
\includegraphics[width=1.0\linewidth]{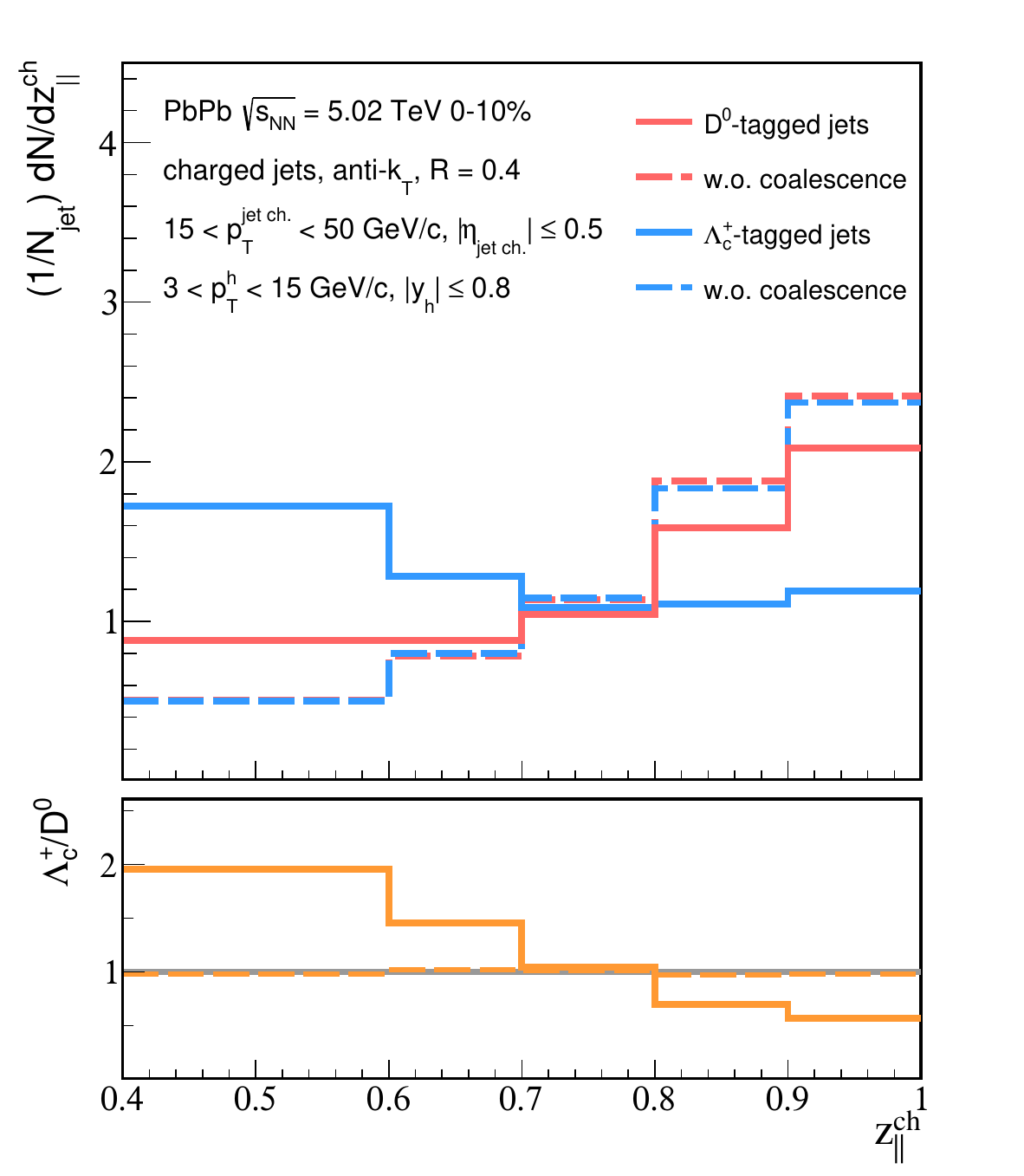}
\caption{ $\Lambda_{c}^{+}$ and $D^{0}$ parallel jet momentum fraction distributions in 0-10\% central $\mathrm{Pb+Pb}$ collisions at $\sqrt{s_{NN}}=5.02$~TeV.
The jets are selected at $|\eta_{\rm jet}|<0.5$ with transverse momentum $7<p_T^{\rm ch\,jet}<15$~GeV/$c$ ($R=0.4$); charmed hadrons are selected in the kinematic range $3<p_T^h<15$~GeV/$c$ and $|y_h|<0.8$.
Solid lines represent the full calculation with coalescence plus fragmentation; dashed lines represent charmed hadrons produced when only fragmentation hadronization is enabled (coalescence turned off).
The ratio of the $z^{ch}_{||}$ distribution for $\Lambda^+_c$ over $D^0$ is shown in the bottom panel. }
\label{fig:zpara-onlyfrag-5020-0_10}
\end{figure}

To disentangle the effect of parton energy loss and hadronization on longitudinal momentum fractions inside jets, we show in Fig.~\ref{fig:zpara-onlyfrag-5020-0_10} the $z^{ch}_{||}$ distributions of $\Lambda_c^+$ and $D^0$ and their ratio in Pb+Pb collisions. The dashed curves are obtained by switching off the coalescence channel and producing charmed hadrons solely via fragmentation, which isolates the effect of parton energy loss. Both the $\Lambda_c^+$ and $D^0$ spectra inside jets rise toward large $z^{ch}_{||}$, and this shape remains consistent with the fragmentation-only baseline in p+p collisions once in-medium energy loss is included.

The solid lines show that the coalescence mechanism redistributes charmed hadrons toward lower $z^{ch}_{||}$, reducing the fraction of the jet momentum they carry. The shift is more pronounced for charmed baryons, so that the $\Lambda_c^+/D^0$ ratio decreases with $z^{ch}_{||}$---a behaviour that is generic to recombination-type hadronization and resembles the color-reconnection case in p+p collisions.

\begin{figure}[tbp]
\centering
\includegraphics[width=1.0\linewidth]{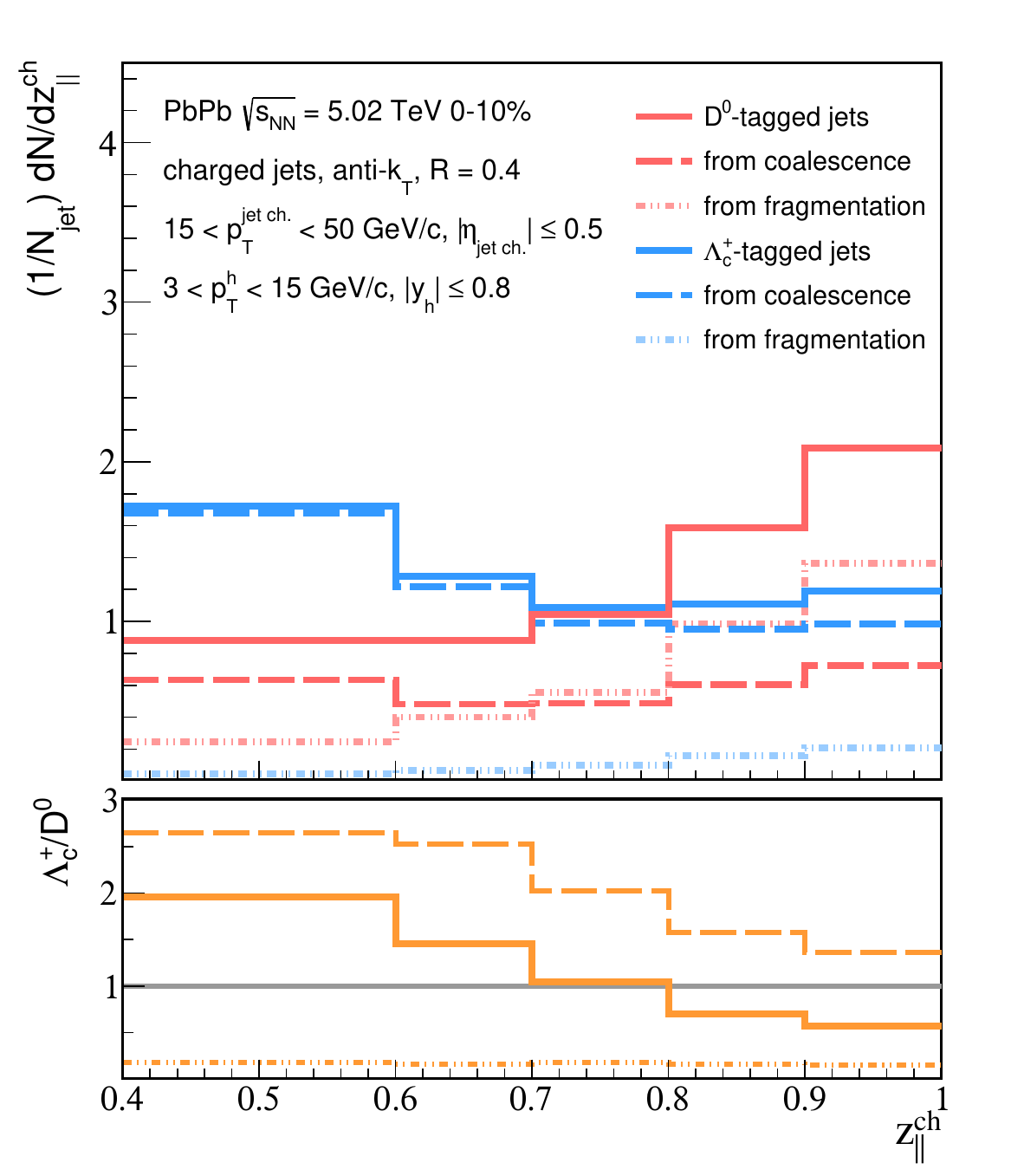}
\caption{ $\Lambda_{c}^{+}$ and $D^{0}$ parallel jet momentum fraction distributions in 0-10\% central $\mathrm{Pb+Pb}$ collisions at $\sqrt{s_{NN}}=5.02$~TeV.
The jets are selected at $|\eta_{\rm jet}|<0.5$ with transverse momentum $7<p_T^{\rm ch\,jet}<15$~GeV/$c$. Charmed hadrons are selected in the kinematic range $3<p_T^h<15$~GeV/$c$ and $|y_h|<0.8$.
Solid lines represent the total spectra; the coalescence and fragmentation contributions are shown as dashed and dash-dotted lines, respectively.
The ratio of the $z^{ch}_{||}$ distribution for $\Lambda^+_c$ over $D^0$ is shown in the bottom panel.}
\label{fig:zpara-from-5020-0_10-LcD0}
\end{figure}

Figure~\ref{fig:zpara-from-5020-0_10-LcD0} shows the $\Lambda_c^+$ and $D^0$ $z^{ch}_{||}$ distributions decomposed into coalescence and fragmentation contributions. In the upper panel, coalescence dominates at small $z^{ch}_{||}$ for both species, decreasing with $z^{ch}_{||}$ for $\Lambda_c^+$ and remaining nearly flat for $D^0$, whereas the fragmentation contribution rises with $z^{ch}_{||}$ for both species and becomes comparable to or larger than the coalescence one at large $z^{ch}_{||}$. Consequently the total $\Lambda_c^+$ distribution decreases with $z^{ch}_{||}$ while the total $D^0$ distribution increases, so that the two cross.
The lower panel reflects this composition: the ratio obtained from the coalescence contribution is the largest and decreases with $z^{ch}_{||}$, that from the fragmentation contribution is the smallest and nearly flat, and the total ratio lies in between and decreases.

\begin{figure}[tbp]
\centering
\includegraphics[width=1.0\linewidth]{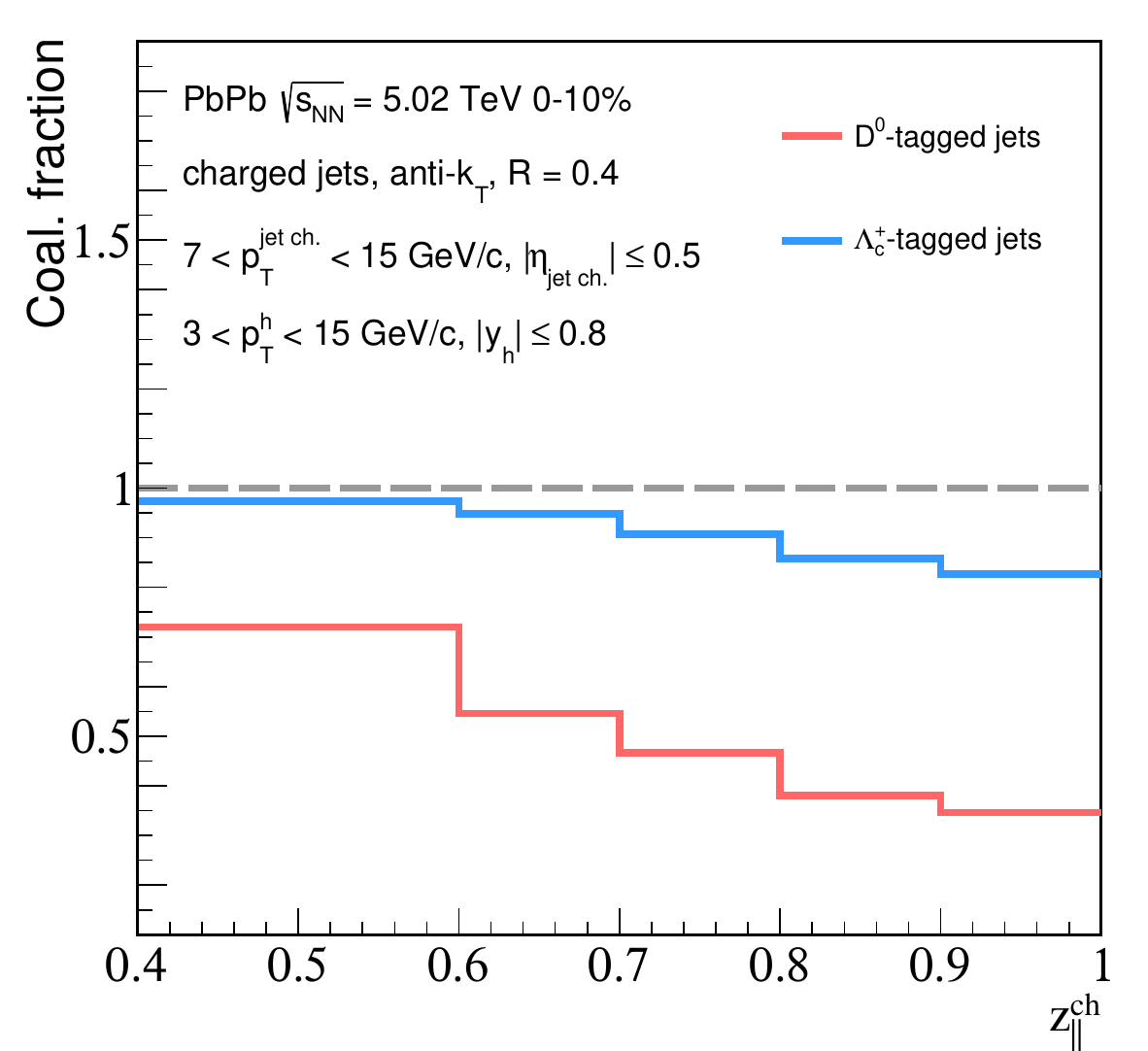}
\caption{ The coalescence fraction for $D^{0}$ and $\Lambda_{c}^{+}$ as a function of their parallel jet momentum fraction in 0-10\% central $\mathrm{Pb+Pb}$ collisions at $\sqrt{s_{NN}}=5.02$~TeV. The jets and charmed hadrons are selected as in Fig.~\ref{fig:zpara-from-5020-0_10-LcD0}. }
\label{fig:zpara-fraction-5020-0_10}
\end{figure}

To understand the $z^{ch}_{||}$ dependence of the $\Lambda_c^+/D^0$ ratio, we examine in Fig.~\ref{fig:zpara-fraction-5020-0_10} the coalescence fraction, defined as the spectrum of hadrons produced via coalescence divided by the total spectrum. For both species the fraction decreases with $z^{ch}_{||}$---from about 0.97 to 0.82 for $\Lambda_c^+$ and from about 0.72 to 0.35 for $D^0$---and it is larger for $\Lambda_c^+$ at all $z^{ch}_{||}$. The larger values for $\Lambda_c^+$ mean that its distribution is the more coalescence-dominated of the two, which is why the total $\Lambda_c^+$ distribution in the upper panel of Fig.~\ref{fig:zpara-from-5020-0_10-LcD0} decreases with $z^{ch}_{||}$ while the total $D^0$ distribution increases. The behaviour of the ratio in the lower panel follows from this composition: since coalescence yields a larger $\Lambda_c^+/D^0$ ratio than fragmentation, and the coalescence fraction of $D^0$ decreases with $z^{ch}_{||}$, the ratio is largest at small $z^{ch}_{||}$, where coalescence dominates, and decreases toward large $z^{ch}_{||}$, where fragmentation becomes important. The decrease of the coalescence fraction reflects the fact that hadrons carrying a larger fraction of the jet momentum are more likely to hadronize via fragmentation.

\begin{figure}[tbp]
\centering
\includegraphics[width=1.0\linewidth]{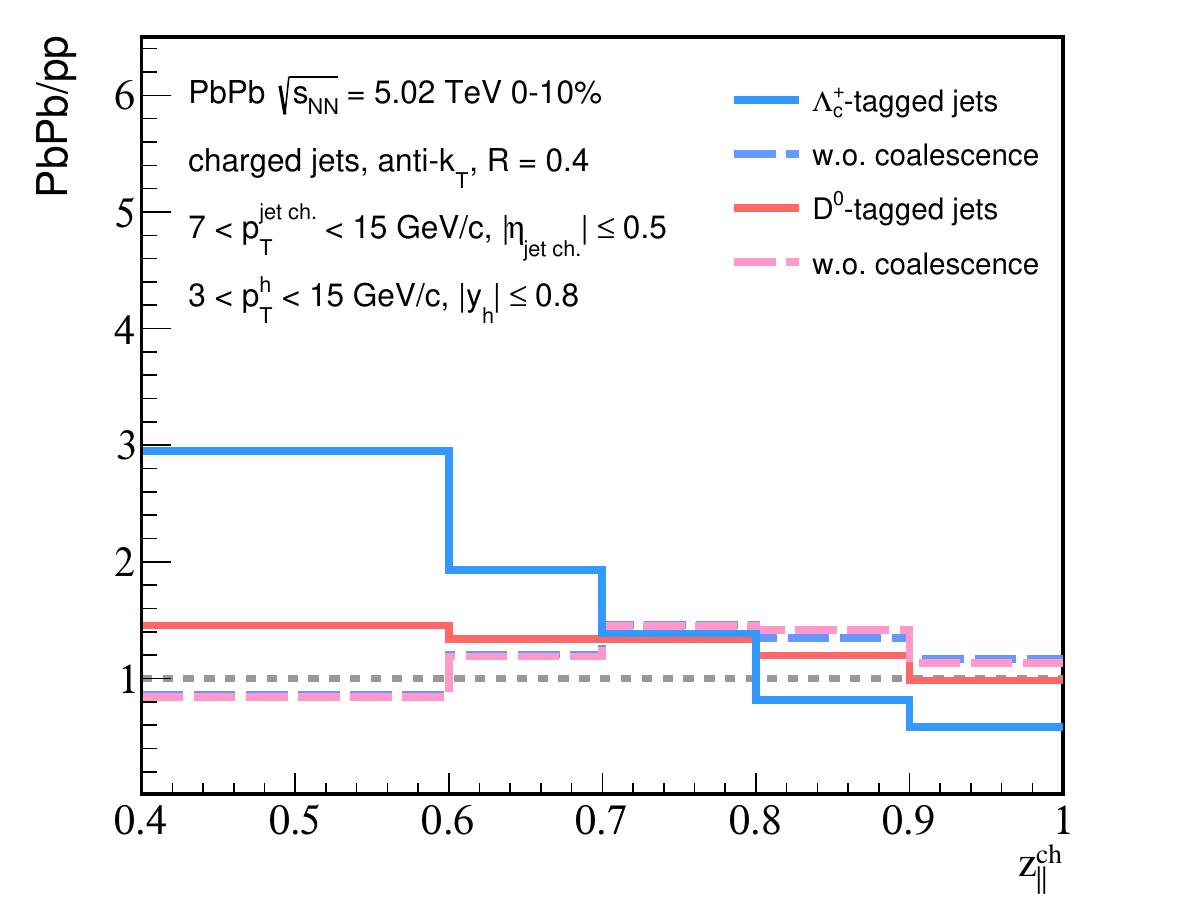}
\includegraphics[width=1.0\linewidth]{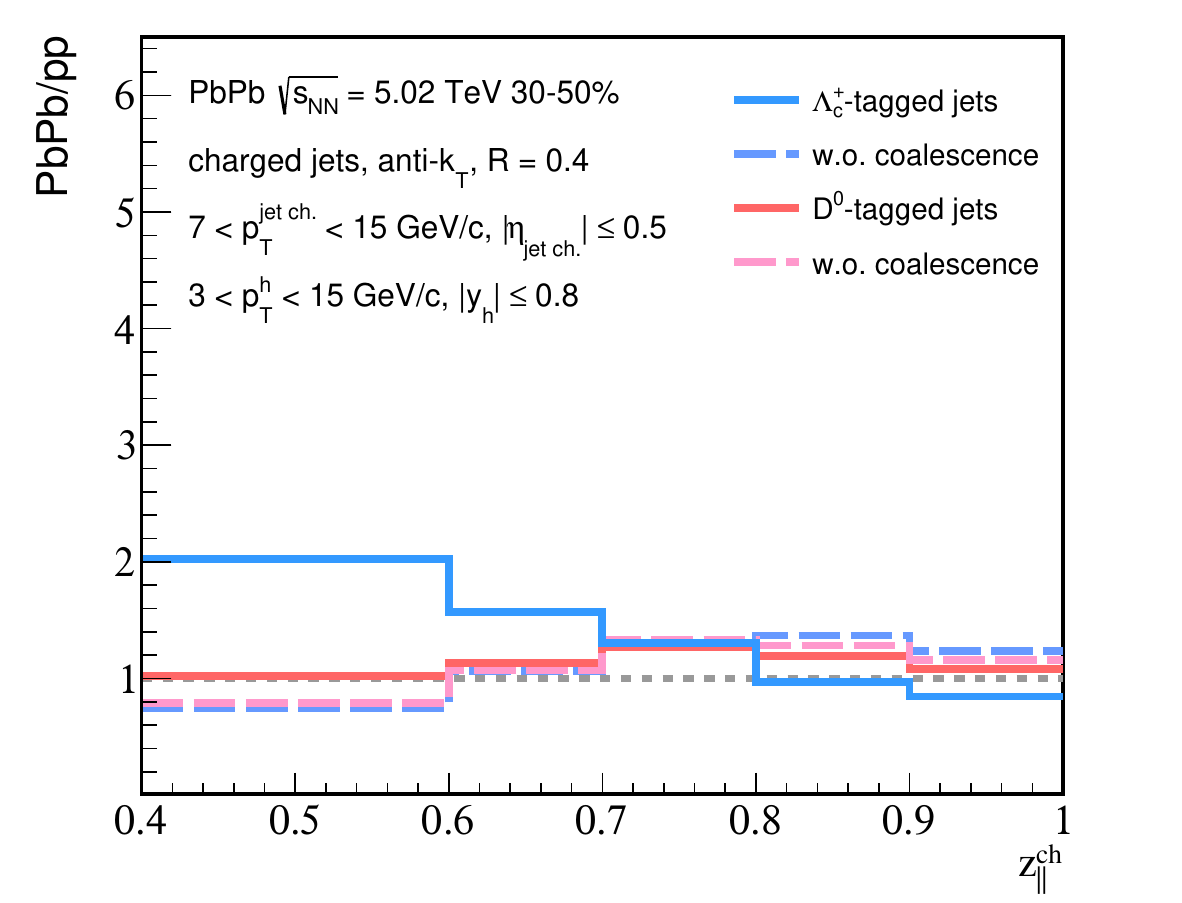}
\caption{
Ratio of the parallel jet momentum fraction $z^{ch}_{||}$ distributions in Pb+Pb to p+p collisions (PbPb/pp) for $\Lambda_{c}^{+}$ and $D^{0}$ in 0-10\% central (upper panel) and 30-50\% central (lower panel) $\mathrm{Pb+Pb}$ collisions at $\sqrt{s_{NN}}=5.02$~TeV, normalized to the fragmentation-function-based p+p baseline (see text). The jets and charmed hadrons are selected as in Fig.~\ref{fig:zpara-from-5020-0_10-LcD0}.
Dashed lines represent charmed hadrons produced when only fragmentation hadronization is enabled (coalescence turned off) in $\mathrm{Pb+Pb}$ collisions. }
\label{fig:zpara-RAA-5020}
\end{figure}

To separate the medium effects on the $z^{ch}_{||}$ spectrum from the vacuum baseline and to explore their centrality dependence, we turn to the ratio of the $z^{ch}_{||}$ distributions in Pb+Pb to those in p+p collisions (denoted PbPb/pp in the figures), shown for $\Lambda_c^+$ and $D^0$ in Fig.~\ref{fig:zpara-RAA-5020}, where the p+p reference is computed with the fragmentation-only scheme. This choice of denominator ensures that numerator and denominator share the same hadronization scheme whenever coalescence is switched off: the PbPb/pp ratio then isolates, by construction, the energy-loss effect (dashed curves), whereas the difference between the solid and dashed curves isolates the coalescence effect on the same event sample. The ratio is therefore a model decomposition tool rather than the experimental nuclear modification factor $R_{AA}$, which would use a measured p+p reference; we return to this point in Sec.~VI. Because both contributions are evaluated on the same event sample, the coalescence effect can be defined in a scheme-independent way as $R_{\rm coal}(z^{ch}_{||})=(dN^{\rm coal+frag}/dz^{ch}_{||})/(dN^{\rm frag\!-\!only}/dz^{ch}_{||})$, which provides the most robust measure of the coalescence mechanism.

The PbPb/pp ratio shows that coalescence shifts more $\Lambda_c^+$ than $D^0$ to small $z^{ch}_{||}$: in 0-10\% central collisions the low-$z^{ch}_{||}$ enhancement reaches about 3 for $\Lambda_c^+$ and about 2 for $D^0$, while in the 30-50\% centrality class it is reduced and the ratio approaches the fragmentation-only behaviour. With coalescence switched off, the distributions are essentially unchanged with respect to p+p collisions in both centrality classes, so that the PbPb/pp ratio is independent of $z^{ch}_{||}$---the shape of the $z^{ch}_{||}$ spectrum is insensitive to parton energy loss in the kinematic region studied. This approximate invariance can be understood because, within the tagged-jet kinematics considered here, energy loss redistributes the momenta of the shower partons without significantly changing the fraction of the jet momentum carried by the leading charmed hadron, so that the modification largely cancels in the PbPb/pp ratio of each species.

\begin{figure}[tbp]
\centering
\includegraphics[width=1.0\linewidth]{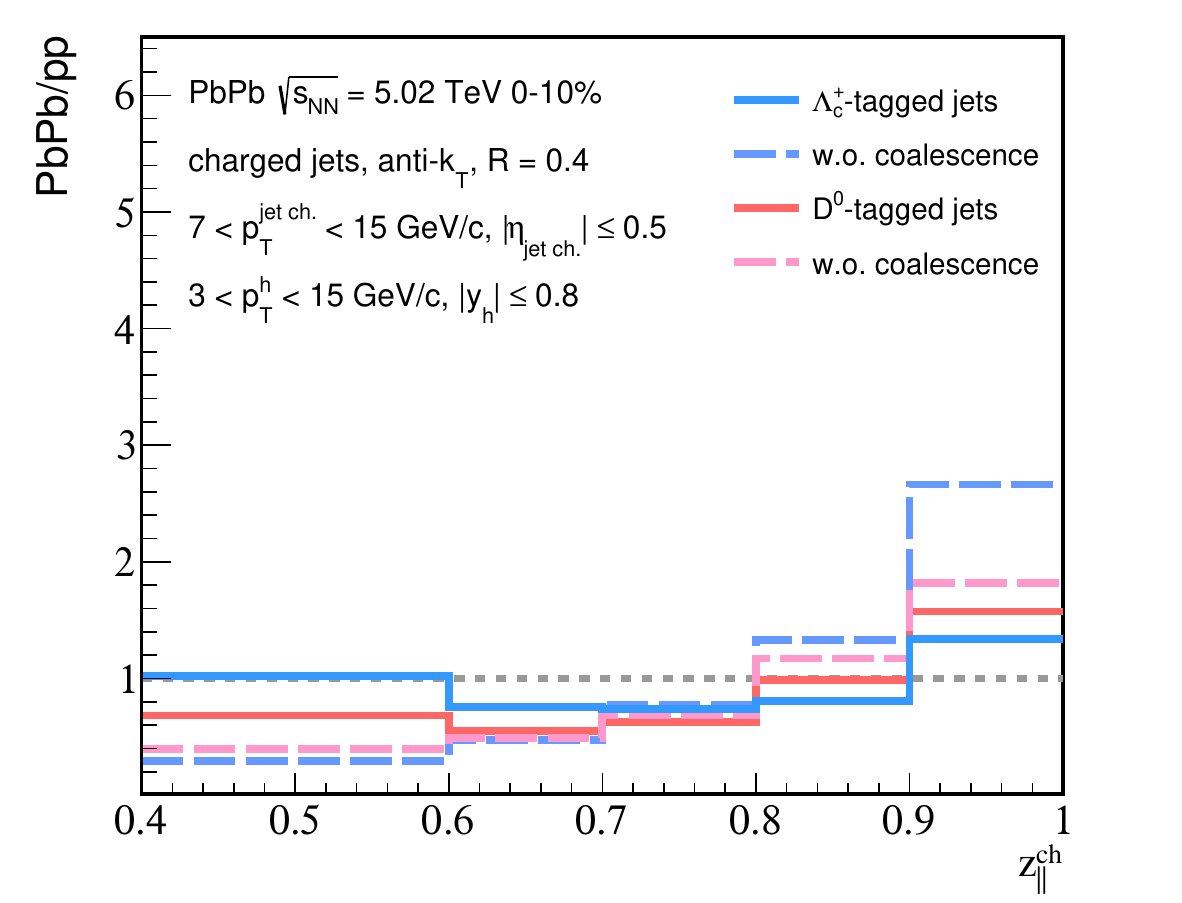}
\includegraphics[width=1.0\linewidth]{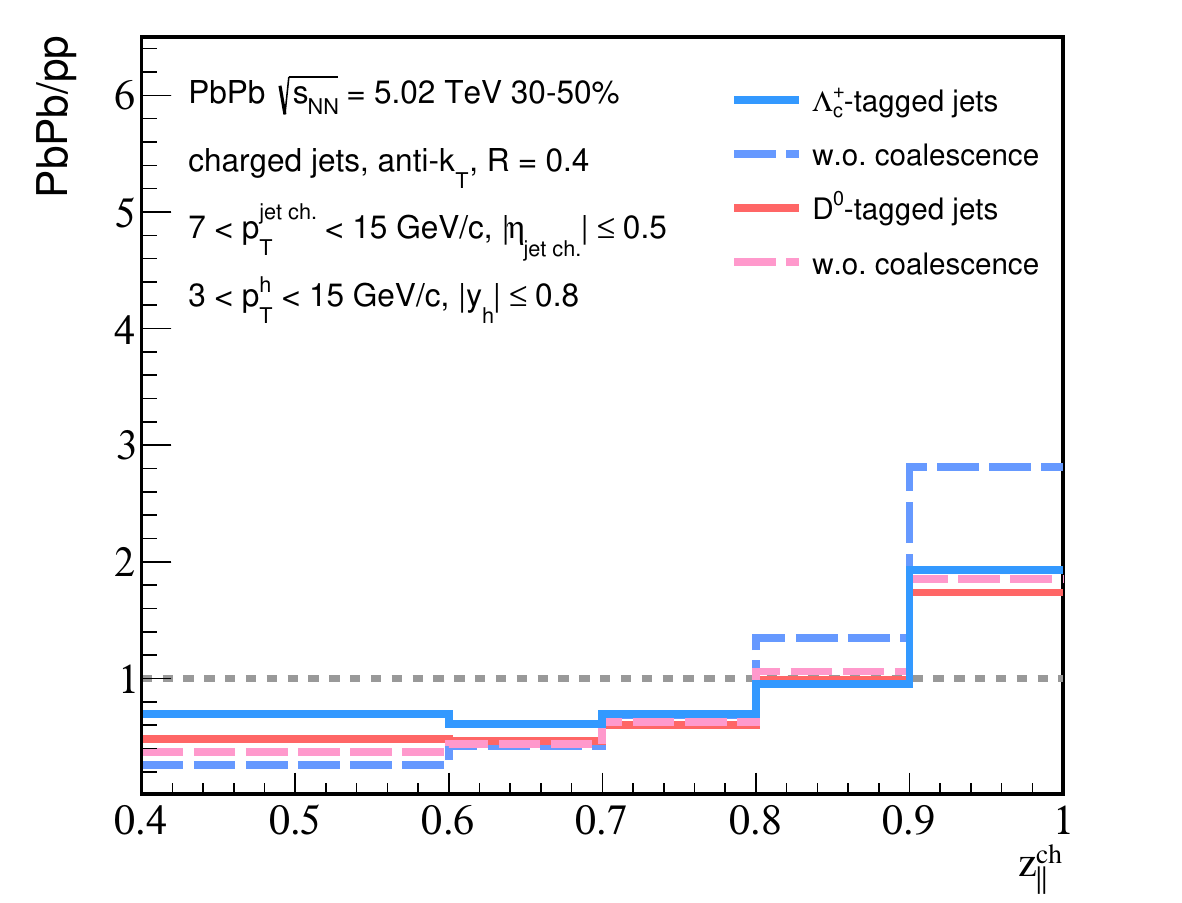}
\caption{
Ratio of the parallel jet momentum fraction $z^{ch}_{||}$ distributions in Pb+Pb to p+p collisions (PbPb/pp) for $\Lambda_{c}^{+}$ and $D^{0}$ in 0-10\% central (upper panel) and 30-50\% central (lower panel) $\mathrm{Pb+Pb}$ collisions at $\sqrt{s_{NN}}=5.02$~TeV, normalized to the CR-BLC Mode 2 p+p baseline (see text). The jets and charmed hadrons are selected as in Fig.~\ref{fig:zpara-from-5020-0_10-LcD0}.
Dashed lines represent charmed hadrons produced when only fragmentation hadronization is enabled (coalescence turned off) in $\mathrm{Pb+Pb}$ collisions. }
\label{fig:zpara-RAA-5020-CRB}
\end{figure}

Because the fragmentation-only p+p baseline differs from the CR-BLC Mode 2 description adopted in Sec.~II, which itself contains a low-$z^{ch}_{||}$ baryon enhancement, we have also computed the PbPb/pp ratio with the CR-BLC Mode 2 p+p reference (Fig.~\ref{fig:zpara-RAA-5020-CRB}). The two references give different $z^{ch}_{||}$ dependences: with the fragmentation-only baseline the ratio is enhanced at small $z^{ch}_{||}$, whereas with the CR-BLC baseline the low-$z^{ch}_{||}$ part is absorbed by the reference and an enhancement appears at large $z^{ch}_{||}$ instead. The size and the $z^{ch}_{||}$ dependence of the PbPb/pp ratio therefore depend on the p+p reference, which confirms that this ratio is a model decomposition tool rather than a measurement of $R_{AA}$, and that a measured p+p reference at 5.02~TeV is required to extract the coalescence effect experimentally.

\section{Discussion and summary}
\subsection{Discussion}
For an experimental counterpart of our PbPb/pp ratio, two caveats should be kept in mind. First, the reference would be the measured p+p $z^{ch}_{||}$ distribution, which is itself better described by beyond-leading-color hadronization scenarios than by fragmentation alone~\cite{ALICE:2023jgm}; as shown in Sec.~V, the resulting PbPb/pp ratio would then differ both in size and in its $z^{ch}_{||}$ dependence from the fragmentation-only one. Second, the existing p+p $z^{ch}_{||}$ measurement was performed at $\sqrt{s}=13$~TeV, whereas a Pb+Pb measurement requires a p+p reference at 5.02~TeV, which is not yet available. Our p+p predictions at 5.02~TeV provide precisely this reference.

We close with a remark on the consistency of the hadronization schemes entering the two sides of our PbPb/pp ratios. The p+p reference is modeled either with fragmentation only or with the CR-BLC Mode~2 tune, whereas the Pb+Pb calculation employs the hybrid coalescence plus fragmentation scheme, so numerator and denominator do not always share a common hadronization scheme. Our ratio is therefore constructed as a decomposition tool. With the fragmentation-only reference, both sides are obtained with the same scheme whenever coalescence is switched off, so that the dashed curves of Fig.~\ref{fig:zpara-RAA-5020} isolate the energy-loss effect, and the solid--dashed difference isolates the coalescence effect on the same Pb+Pb event sample. The latter is defined in a scheme-independent way by $R_{\rm coal}$, which involves Pb+Pb quantities only.

For the CR-BLC Mode~2 reference, the dashed curves no longer isolate the energy-loss effect alone, since numerator and denominator then differ in their hadronization scheme as well; this is why the size and the $z^{ch}_{||}$ dependence of the PbPb/pp ratio depend on the reference, whereas $R_{\rm coal}$ does not. An experimental measurement with the actual p+p reference at 5.02~TeV would remove this model dependence altogether.

A related question concerns a possible small-system medium in p+p collisions. If a hot droplet were formed in high-multiplicity p+p events, charmed baryons would also be assembled by coalescence there, raising the inclusive $\Lambda_c^+/D^0$ ratio, while the in-jet ratio should be much less affected, because the coalescence partners would be thermal partons uncorrelated with the jet and would be preferentially discarded by the jet selection. Our Pb+Pb results are consistent with this expectation: coalescence suppresses the in-jet ratio only moderately, whereas the inclusive $\Lambda_c^+/D^0$ ratio is enhanced by a much larger factor; in a much smaller p+p droplet the residual in-jet effect would be expected to be smaller still. A dedicated small-system implementation of the hybrid hadronization, e.g., a core-corona approach in multiplicity-selected p+p collisions, would allow a direct test of this scenario.

Several extensions of this work are possible. On the p+p side, multiplicity-dependent in-jet measurements would probe the small-system-medium scenario discussed above. On the heavy-flavor side, the same jet-tagged strategy can be applied to the $\Xi_c$ and $\Omega_c$ baryons, whose coalescence component is expected to be even more pronounced, and to jets tagged by other charmed mesons such as $D^+$ and $D_s^+$, for which experimental measurements may become available sooner. A measurement of the in-jet $\Lambda_c^+/D^0$ ratio in Pb+Pb collisions would provide a direct test of the predictions presented here.

\subsection{Summary}
We have investigated the in-jet $\Lambda_c^+/D^0$ ratio and the parallel jet momentum fraction distributions of $\Lambda_{c}^{+}$ and $D^{0}$ in p+p and Pb+Pb collisions, using PYTHIA8 for p+p and the SHELL model with a coalescence plus fragmentation hybrid hadronization for Pb+Pb, with the aim of discriminating between candidate hadronization mechanisms. The in-jet ratio behaves oppositely under the two mechanisms that enhance the inclusive ratio: color reconnection in p+p enhances the in-jet $\Lambda_{c}^{+}/D^0$ ratio at low-to-intermediate $p_T$, with an enhancement that grows with the jet $p_T$ cut, whereas in-medium coalescence in Pb+Pb suppresses the in-jet ratio moderately at intermediate $p_T$. We trace this sign flip to the different correlation of the produced hadrons with the jet: the low-$p_T$ color-reconnection products remain correlated with the jet and are displaced towards larger distances from the axis, while coalescence hadrons pick up momentum from soft thermal partons uncorrelated with the jet direction and are preferentially discarded by the jet selection, driving the in-jet ratio toward its fragmentation-only value. Both mechanisms shift charmed hadrons toward small $z^{ch}_{||}$, with a stronger effect on $\Lambda_c^+$ than on $D^0$, so that the $z^{ch}_{||}$ spectrum is a clean probe of recombination-type hadronization whose shape is insensitive to parton energy loss when compared with the fragmentation-only p+p reference; the identity of the mechanism is fixed by combining the $z^{ch}_{||}$ spectrum with the sign of the in-jet $p_T$ ratio and with the radial profile of the enhancement, which increases with the distance from the jet axis in both collision systems, the interval displaced farthest outward being the lowest-$p_T$ one in p+p and the intermediate-$p_T$ one in Pb+Pb.

Because the coalescence fraction inside jets differs between the two species---it is larger for $\Lambda_c^+$ than for $D^0$ and decreases with $z^{ch}_{||}$---in-jet observables constrain the hybrid hadronization in a way that cannot be obtained from inclusive measurements alone. Our p+p results at $\sqrt{s}=5.02$~TeV provide the reference baseline required for the corresponding jet-tagged measurements in Pb+Pb collisions.

\section{ACKNOWLEDGMENTS}
B.W.Z. is supported by a grant from the National Natural Science Foundation of China (Key Program) (No.~12535010).

\bibliography{xuref}

\end{document}